 \documentclass[twocolumn,tighten]{aastex62}
\newcommand{\atl}{ATLAS$^{\rm 3D}$}

\newcommand{\siglos}{$\sigma_{\rm los}(R)$}

\usepackage{amsmath}

\shorttitle{Radial Acceleration Relation in the Super-critical Regime}
\shortauthors{Chae et al.}

\begin{document}

\title{Radial Acceleration Relation between Baryons and Dark or Phantom Matter \\in the Super-critical Acceleration Regime of Nearly Spherical Galaxies}

\author{Kyu-Hyun Chae}
\affil{Department of Physics and Astronomy, Sejong University, 
 209 Neungdong-ro Gwangjin-gu, Seoul 05006, Republic of Korea}
\affil{Graduate Program of Astronomy and Space Science, Sejong University, 209 Neungdong-ro Gwangjin-gu, Seoul 05006, Republic of Korea}
\email{KHC: chae@sejong.ac.kr, kyuhyunchae@gmail.com}

\author{Mariangela Bernardi}
\affil{Department of Physics and Astronomy, University of Pennsylvania, 209 South 33rd Street, Philadelphia, PA 19104, USA}
\email{MB: bernardm@physics.upenn.edu}

\author{Ravi K. Sheth}
\affil{Department of Physics and Astronomy, University of Pennsylvania, 209 South 33rd Street, Philadelphia, PA 19104, USA}
\email{RKS: shethrk@physics.upenn.edu}

\author{In-Taek Gong}
\affil{Graduate Program of Astronomy and Space Science, Sejong University, 209 Neungdong-ro Gwangjin-gu, Seoul 05006, Republic of Korea}



\begin{abstract}
  The central regions of nearby elliptical galaxies are dominated by baryons (stars) and provide interesting laboratories for studying the radial acceleration relation (RAR).  We carry out exploratory analyses and discuss the possibility of constraining the RAR in the super-critical acceleration range $(10^{-9.5},\hspace{1ex}10^{-8})$~${\rm m}~{\rm s}^{-2}$ using a sample of nearly round pure-bulge (spheroidal, dispersion-dominated) galaxies including 24 ATLAS$^{\rm 3D}$ galaxies and 4201 SDSS galaxies covering a wide range of masses, sizes and luminosity density profiles. We consider a range of current possibilities for the stellar mass-to-light ratio ($M_\star/L$), its gradient and dark or phantom matter (DM/PM) halo profiles. We obtain the probability density functions (PDFs) of the parameters of the considered models via Bayesian inference based on spherical Jeans Monte Carlo modeling of the observed velocity dispersions. We then constrain the DM/PM-to-baryon acceleration ratio $a_{\rm X}/a_{\rm B}$ from the PDFs.  Unless we ignore observed radial gradients in $M_\star/L$, or assume unreasonably strong gradients, marginalization over nuisance factors suggests $a_{\rm X}/a_{\rm B} = 10^{p} (a_{\rm B}/a_{+1})^q$ with $p = -1.00 \pm 0.03$ (stat) $^{+0.11}_{-0.06}$ (sys) and $q=-1.02 \pm 0.09$ (stat) $^{+0.16}_{-0.00}$ (sys) around a super-critical acceleration $a_{+1}\equiv 1.2\times 10^{-9}~{\rm m}~{\rm s}^{-2}$. In the context of the $\Lambda$CDM paradigm, this RAR suggests that the NFW DM halo profile is a reasonable description of galactic halos even after the processes of galaxy formation and evolution. In the context of the MOND paradigm, this RAR favors the Simple interpolating function but is inconsistent with the vast majority of other theoretical proposals and fitting functions motivated mainly from sub-critical acceleration data.
\end{abstract}

\keywords{ dark matter --- gravitation --- galaxies: kinematics and dynamics -- galaxies: structure}



\section{Introduction} \label{sec:intro}

Mass discrepancy in galaxies -- the disagreement between the mass inferred from the observed light distribution and that derived from kinematics of stars and gases or gravitational lensing under the standard Newton-Einstein gravity and dynamics -- remains an unsolved problem for fundamental physics and cosmology. Proposed resolutions of the mystery broadly fall into two classes:  much of the gravitating mass is dark matter (DM), or our current understanding of gravity or dynamics must be modified (MG) (e.g.\ \citealt{FM,BT18}).

Recent decades have witnessed the surprising observation that the mass discrepancy becomes prominent only when the (centripetal) radial acceleration is weaker than a critical value $a_0 \sim 10^{-10}$~m~s$^{-2}$.  In this regime, the observed radial acceleration traced by the circular velocity $V$, $a = V^2 /r$, is well correlated with the Newtonian radial acceleration predicted by the distribution of baryons, $a_{\rm B}$ (e.g.\ \citealt{McG04,MLS,Lel}). The radial acceleration relation (RAR, or mass discrepancy acceleration relation) is now well-documented for rotating galaxies, particularly in the low (sub-critical) acceleration regime $a_{\rm B} < 10^{-10}$~m~s$^{-2}$ \citep{MLS,Lel,Li18}.

The empirical RAR has recently prompted hydrodynamic-simulation (e.g.\ \citealt{Lud,Ten,KW}) and semi-analytic (e.g.\ \citealt{Des,Nav}) studies in the standard Lambda cold dark matter ($\Lambda$CDM) paradigm. In the $\Lambda$CDM paradigm the empirical RAR simply represents an average property of galaxies which is a consequence of galactic astrophysics: as such, there is no need for all galaxies to follow the same RAR. However, it is argued that $\Lambda$CDM has difficulty in reproducing \citep{vP18}, or can only qualitatively reproduce \citep{Lel,Ten}, the empirical RAR. Whether the empirical RAR arises naturally in the $\Lambda$CDM paradigm and whether its approximate universality indicates a deeper Kepler-like law of galactic dynamics (see \citealt{Rod18,McG18,Kro18}) are currently open questions.

For spherical gravitating systems, the empirical RAR $a=a(a_{\rm B})$ [or $a_{\rm B}=a_{\rm B}(a)$] can be written as
\begin{equation}
a(r) = G \frac{M_{\rm B}(r)+M_{\rm DM}(r)}{r^2} = \left(1+\frac{a_{\rm DM}(r)}{a_{\rm B}(r)}\right) a_{\rm B}(r) 
 \label{eq:RARdm}
\end{equation}
assuming DM, or
\begin{equation}
a(r) = f \left(\frac{a_{\rm B}(r)}{a_0}\right) a_{\rm B}(r)\hspace{1ex}
 \label{eq:RARmd}
\end{equation}
assuming modified Newtonian dynamics (MOND) \citep{Mil} or MG. In Equation~(\ref{eq:RARdm}), $G$ is Newton's gravitational constant and $M_i(r)$ is the mass within the spherical radius $r$ for the $i$th-component. In Equation~(\ref{eq:RARmd}) $f(x)$ is a fitting function, known as the interpolating function (IF). These approaches can be parameterized in a unified way by introducing a `phantom' matter (PM) in MG as follows:
\begin{equation}
  \frac{a}{a_{\rm B}}-1 = f\left(\frac{a_{\rm B}}{a_0} \right)-1
                      \equiv \frac{a_{\rm X}}{a_{\rm B}},
  \label{eq:XM}
  \end{equation}
where X denotes DM or PM.  Working with $a_{\rm X}$ hides the fact that, in $\Lambda$CDM, one treats the discrepancy between what the light predicts and what is observed by adding an independent term, $M_{\rm DM}(<r)$ to the total mass.
In MOND, the amount to be `added' depends on what is present.

\begin{figure}
\begin{center}
\setlength{\unitlength}{1cm}
\begin{picture}(10,10)(0,0)
\put(-1.2,-1.5){\includegraphics{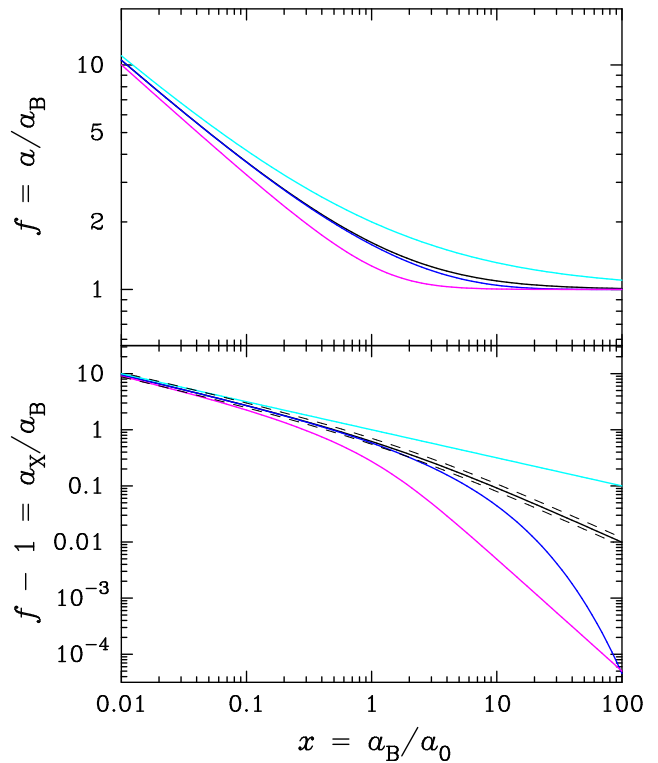}}
\end{picture}
\caption{Various fitting (interpolating) functions of the RAR in the form of $f=a/a_{\rm B}$ (top) or $f-1=a_{\rm X}/a_{\rm B}$ (bottom): (1) black: simple IF (Equation~(\ref{eq:IFnu}), $\nu=1$), (2) magenta: standard IF (Equation~(\ref{eq:IFnu}), $\nu=2$), (3) blue: McGaugh's IF (Equation~(\ref{eq:IFlam}), $\lambda=1$), (4) cyan: Bekenstein's IF (Equation~(\ref{eq:IFgam}), $\gamma=1/2$). The thin black dashed curves in the bottom panel indicate the effects of varying $a_0$ from a fiducial value of 1.2 (e.g., in units of $10^{-10}$~m~s$^{-2}$) to 1.0 and 1.4.}
\label{IFs}
\end{center}
\end{figure}

Several functional forms (see Figure~\ref{IFs}), either empirically-motivated or theory-inspired, have been suggested for the RAR.  These include the `simple' function \citep{FB}, the `standard' function \citep{Ken}, McGaugh's function \citep{McG08}, and Bekenstein's function \citep{Bek}. Theoretical proposals range from Bekenstein's modified gravity \citep{Bek}, Verlinde's emergent gravity \citep{Ver}, a symmetron-like fifth force \citep{BCM} and modified properties of DM such as superfluidity \citep{Kho} or dipolarity \citep{BL}. 

We will consider three families of models under the MOND paradigm. First, 
\begin{equation}
  f_\nu(x) = \left(\frac{1}{2}+\sqrt{\frac{1}{4}+\frac{1}{x^{\nu}}}\right)^{1/\nu}, 
 \label{eq:IFnu}
\end{equation}
with $0<\nu\le 2$; this includes the simple ($\nu=1$) and standard ($\nu=2$) IFs as special cases.  Second, 
\begin{equation}
  f_\lambda(x) = \frac{1}{\left(1-{\rm e}^{-x^{\lambda/2}}\right)^{1/\lambda}} ,
 \label{eq:IFlam}
\end{equation}
with $0.3<\lambda<1.7$, which includes McGaugh's IF ($\lambda=1$) as a special case. Third,
\begin{equation}
  f_\gamma(x) = 1+\frac{1}{x^{\gamma}} ,
 \label{eq:IFgam}
\end{equation}
with $0<\gamma<2$, which includes Bekenstein's IF ($\gamma=1/2$) as a special case.

In MOND, both the IF shape (the value of $\nu$, $\lambda$ or $\gamma$) and the scale $a_0$ are assumed to be the same for all galaxies.  While there is general agreement that $a_0\approx 1.2\times 10^{-10}$~m~s$^{-2}$, there is less agreement on the shape.  Although several previous RAR studies favored the simple IF (e.g., \citealt{FB,SN07,Mil12,CG15,Jan16}), a recent series of analyses \citep{MLS,Lel,Li18} argue that all galaxies follow McGaugh's IF.  This has far-reaching implications:  Any theory of DM or MG that does not predict the exponential decay of $a_{\rm X}/a_{\rm B}$ in the high acceleration limit would be ruled out. For example, a recent proposal of dark matter-baryon interactions \citep{FKP} would be invalidated by McGaugh's function.

However, the top panel of Figure~\ref{IFs} shows that, if one only considers the ratio $a/a_{\rm B}$, which has been the focus of RAR studies to date, then there is only a subtle difference between McGaugh's IF and the simple IF.  On the other hand, as the bottom panel shows, the quantity $a_{\rm X}/a_{\rm B}$ provides greater discriminatory power, especially in the super-critical ($a_0> 1.2\times 10^{-10}$~m~s$^{-2}$) regime.  (Note that changing $a_0$ merely shifts curves left or right; reasonable $\sim 20\%$ changes in $a_0$ cannot bring one IF into agreement with another.  This will simplify some of our analysis below.)  Since the existence and form of the RAR has fundamental implications, in this work we explore the super-critical acceleration regime.  

In this work, we use a Bayesian inference of the ratio $a_{\rm X}/a_{\rm B}$ (Figure~\ref{IFs}), based on Jeans dynamical analyses of nearly spherical galaxies. There are several reasons for using spherical galaxies. First, the optical regions within the effective radius $R_{\rm e}$, which is defined as the two-dimensional radius (projected on the plane of the sky) containing one half of the total light, cover the super-critical acceleration range $10^{-10}{\rm m}~{\rm s}^{-2} \lesssim a_{\rm B} \lesssim 10^{-8}{\rm m}~{\rm s}^{-2}$. Second, they can be described by spherical models avoiding the uncertainties related to angular dependences. Third, spherical galaxies are dispersion (pressure)-supported systems dynamically distinct from rotationally-supported systems, and hence they provide an independent probe of the RAR. Indeed, there exists previous work arguing that the RAR for ellipticals differs from that for disk galaxies (e.g.\ \citealt{Ger01,Jan16}). Finally, spherical systems allow most straightforward tests of theories. In particular, Verlinde's emergent gravity has a specific prediction only for spherical systems. 

To constrain $a_{\rm X}/a_{\rm B}$ in the super-critical regime using elliptical galaxies, a number of factors have to be dealt with carefully. In particular, radial gradients in the stellar mass-to-light ratio $\Upsilon_\star\equiv M_\star/L$ in the central regions of elliptical galaxies (e.g.\ \citealt{MN,LaB16,vD17,Sar18,Old18,Son18}) can affect the inferred $a_{\rm X}/a_{\rm B}$ \citep{Ber18}. The effects of these gradients have been ignored in previous RAR analyses, so in this work we will pay particular attention to them, following the methodology of \citet{CBS18}.

We carry out our analyses in both the $\Lambda$CDM and MOND paradigms, because the empirical RAR has different implications for the two paradigms. In $\Lambda$CDM, a universal RAR is not required by theory; hence, by constraining RARs for individual galaxies and then averaging over the population, one obtains a mean RAR that may be useful for constraining the gastrophysics of galaxies. However, in the case of MOND, a single RAR is supposed to apply universally for {\it all} galaxies; in this case, the theory can be falsified if one can demonstrate that there are statistically significant variations across the population (e.g. from one elliptical galaxy to another) or between populations (e.g. ellipticals vs spirals).

This paper is structured as follows. In Section~\ref{sec:data}, we describe the data and the $\Lambda$CDM and MOND models we consider.  In Section~\ref{sec:lcdm} we apply our Bayesian methodology to infer parameters in the $\Lambda$CDM paradigm. Specifically,  we first describe our Monte Carlo (MC) sampling and Bayesian inference methods in Section~\ref{sec:bayes}, and then derive RARs of individual galaxies where dynamical information over a range of scales is available in Section~\ref{sec:indRAR}.  We discuss what these individual RARs imply for MOND in Section~\ref{sec:lcdm2mond}. In Section~\ref{sec:RARsdss} we modify our methods to analyze galaxies where information from only a single scale is available.  In Section~\ref{sec:stackRAR} we derive an average, empirical RAR by stacking together all the individual results and compare it with a number of MOND predictions.  In Section~\ref{sec:mond} we provide a similar analysis, but now explicitly under the MOND paradigm.  Section~\ref{sec:univ} discusses the issue of universality, and Section~\ref{sec:chi2} provides direct $\chi^2$ tests of MOND models using the observed velocity dispersion profiles of the most spherical galaxies. In Section~\ref{sec:sys}, we discuss potential systematic errors and why we do not think they have biased our results. Finally, in Section~\ref{sec:dis}, we discuss the implications of our exploratory results for DM or MG and future prospects of using integral field spectroscopy (IFS) data on elliptical galaxies for the DM/MG problem. Appendix~\ref{sec:mc2mc} discusses MC sampling methods and their effects on our Bayesian inferences. Appendix~\ref{sec:corner} presents examples of full parameter correlations from our modeling results of the four roundest galaxies in our sample. Throughout we use the following cosmological parameters whenever working in the $\Lambda$CDM framework: $\Omega_{\rm m 0}=0.3$, $\Omega_{\Lambda 0}=0.7$, and $h=H_0 /100$~km~s$^{-1}$~Mpc$^{-1}$ $=0.7$.

\section{Data and Method} \label{sec:data}

\subsection{Datasets} 
We use nearly round pure-bulge (i.e.\ pure ellipsoids without detectable disks) galaxies selected from the \atl\ project \citep{Cap11} and the SDSS DR7 \citep{DR7}; see \citet{CBS18,CBS19} for sample-selection details.  Of the 24 \atl\ galaxies we select (from a total of 260), 16 are kinematic slow rotators (SRs).   We require SDSS galaxies to be statistically similar to the \atl\ galaxies, and so select 4201 galaxies from the UPenn spectroscopic catalog of about 0.7~million galaxies \citep{Mee}. 

In these galaxies, individual stellar velocities and orbits are not observed, but the rms scatter of the line-of-sight velocity component, referred to as the velocity dispersion, is observed through Doppler broadening of spectral lines.  However, \atl\ provides more information about the velocity dispersion than does SDSS.  Specifically, for each \atl\ galaxy, a 2-dimensional map of line-of-sight velocity dispersions is available, from the central region out to (approximately) the projected half-light radius $R_{\rm e}$.  From these we construct $\sigma_{\rm los}(R)$, the line-of-sight velocity dispersion profile \citep{CBS18}. (Throughout this paper, a capitalized $R$ refers to a scale projected onto the plane of the sky.) We also have a measured surface brightness profile $I(R)$ out to a few $R_{\rm e}$, which can be deprojected to give the volume density profile of luminosity $\rho_{\rm L}(r)$ (we use the lower case $r$ to indicate that it is a three-dimensional rather than projected quantity).

For SDSS galaxies, the line-of-sight velocity dispersion is only measured within a single aperture of radius $R_{\rm ap}=D\times \theta_{\rm ap}$, where $\theta_{\rm ap}=1.5$ arcsec and $D$ is the angular-size distance to the galaxy. This quantity is related to the hidden profile by
\begin{equation}
 \sigma_{\rm ap}\equiv \langle \sigma_{\rm los}\rangle (R_{\rm ap}) =
  \frac{ \int_0^{R} I(R')\, \sigma_{\rm los}(R')\, R' dR'}{\int_0^{R} I(R')\,  R' dR'},
  \label{eq:VDap}
\end{equation}
where $\sigma_{\rm los}(R)$ is the velocity dispersion profile and $I(R)$ is the surface brightness distribution. Therefore, as we describe below, our treatment of SDSS galaxies will be slightly different than for ATLAS$^{\rm 3D}$.  

\subsection{Relation to Jeans equation}\label{sec:Jeans}
The line-of-sight velocity dispersion at projected radius $R$ on the sky is related to the three-dimension dispersion by 
\begin{equation}
 \sigma_{\rm los}^2(R)=\frac{2}{I(R)} \int_{R}^{\infty}
 \rho_{\rm L}(r) \sigma_{\rm r}^2(r) \left[ 1 - \frac{R^2}{r^2} \beta(r) \right]
 \frac{r dr}{\sqrt{r^2-R^2}},
\label{eq:losvd}
\end{equation}
where $\sigma_{\rm r}^2(r)$ is the radial velocity dispersion and
$\beta(r)\equiv 1 - \sigma_{\rm t}^2(r)/\sigma_{\rm r}^2(r)$,
where $\sigma_{\rm t}^2(r)\equiv \left[\sigma_{\theta}^2(r)+\sigma_{\phi}^2(r)\right]/2$ is the velocity dispersion in the tangential (i.e.\ angular in the spherical polar coordinates) direction, is the velocity dispersion anisotropy.

In principle, the spherical Jeans equation (Equation~4.215 of \citealt{BT}) -- which is satisfied if spherical galaxies are in equilibrium -- allows the observed line-of-sight velocity dispersions to constrain the acceleration $a(r)$, where $a$ is given by either Equation~(\ref{eq:RARdm}) ($\Lambda$CDM) or Equation~(\ref{eq:RARmd}) (MOND).  This is because the spherical Jeans equation relates $\sigma_{\rm r}^2(r)$ to $a(r)$: 
 \begin{equation}
 \frac{d[\rho_{\rm B}(r) \sigma_{\rm r}^2(r)]}{dr} 
  + 2 \frac{\beta(r)}{r} [\rho_{\rm B}(r) \sigma_{\rm r}^2(r)]
  = - \rho_{\rm B}(r) a(r),
 \label{eq:Jeans}
\end{equation}
 where $\rho_{\rm B}(r)$ is the density profile of the baryons.  For our pure-bulge systems, we assume this is the same as that of the stellar mass, so that $\rho_{\rm B}(r) \equiv \Upsilon_\star(r)\,\rho_{\rm L}(r)$, where $\rho_{\rm L}(r)$ is the deprojected light profile, and $\Upsilon_\star\equiv M_\star/L$ is the stellar mass-to-light ratio.  Note that $a(r) = a_{\rm B}(r) + a_{\rm X}(r)$, where $a_{\rm B}$ depends on the baryonic profile $\rho_{\rm B}(r)$ whereas $a_{\rm X}(r)$ depends on the DM or PM profile $\rho_{\rm X}(r)$ (Equation~(\ref{eq:XM})).

 Thus, in practice, for each ATLAS$^{\rm 3D}$ galaxy, the two observed profiles $\sigma_{\rm los}(R)$ and $I(R)$ constrain three unknown functions of scale: $\Upsilon_\star(R)$,  $\rho_{\rm X}(r)$, and $\beta(r)$. (Strictly speaking, the unknown is $\Upsilon_\star(r)$, but, as will become clear shortly, it is more convenient to consider the projected profile instead.)  Therefore, our approach is to adopt well-motivated, flexibly parameterized models for the three unknown profiles.
 
\subsection{Model parameterization}
For the projected stellar mass-to-light profile we set
\begin{equation}
  \Upsilon_\star (R) =\Upsilon_{\star 0} \times \max\left\{1+ K\left[A - B (R/R_{\rm e})\right],1\right\},
 \label{eq:MLgrad}
\end{equation}
with $A=2.33$ and $B=6$ \citep{Ber18,CBS18}.
(We show our results are robust to reasonable changes in these values in Section~\ref{sec:sys}.)
The `gradient strength' parameter $K$ accounts for recent evidence of radial gradients in the central regions ($<0.4 R_{\rm e}$) (e.g.\ \citealt{vD17,MN,Son18}). We allow $0 \le K < 1.5$ as this includes fully the range of gradient strengths reported in the literature.  

For the VD anisotropy we use a generalized Osipkov-Merritt (gOM) model:
\begin{equation}
  \beta_{\rm gOM}(r)=\beta_0 + (\beta_\infty - \beta_0) \frac{(r/r_a)^2}{1+(r/r_a)^2},
 \label{eq:gOM}
\end{equation}
(\citet{BT}, p. 297), with allowed prior ranges $-2 < \beta_{\rm gOM}(r) < 0.7$ for all $r$ and $0<r_a< R_{\rm e}$ \citep[following][]{CBS18}.

Finally, we must specify a model for $a_{\rm X}(r)$.  In $\Lambda$CDM, we compute it by setting $\rho_{\rm X}$ equal to 
\begin{equation}
  \rho_{\rm gNFW}(r) \propto
  r^{-\alpha} \left[1+c_{200}\left(\frac{r}{r_{200}}\right)\right]^{-3+\alpha}.
 \label{eq:gNFW} 
\end{equation}
This is known as a generalized \cite{NFW} model (hereafter gNFW).  Here $r_{200}$ is the radius of the sphere within which the DM density is 200 times the cosmic mean matter density, $\alpha$ is the inner slope, and the outer slope is $3$.  Integrating this over a sphere of radius $r$ yields $M_{\rm DM}(r)$ and hence $a_{\rm DM}(r) = GM_{\rm DM}(r)/r^2$. We allow $0.1< \alpha < 1.8$ (the NFW value is $\alpha_{\rm NFW}=1$). The halo mass $M_{200}\equiv M_{\rm DM}(r_{200})$ is constrained using the weak-lensing derived $M_\star^{\rm Krou}$-$M_{200}$ relation given in Table~\ref{tab:MsM200} \cite[taken from][]{Man16}, where $M_\star^{\rm Krou}$ is the stellar mass derived assuming the Kroupa IMF \citep{Kro02}. The concentration parameter $c_{200}$ is, in principle, free but we impose the constraint \citep{Man08,Man16} that the outer profile $r > 0.2 r_{200}$ mimics the NFW profile seen in N-body simulations. Then, following Section~3.3.1 of \citet{CBK14}, we set \begin{equation}
  c_{200}=\max\left[
    \left(\frac{3-\alpha}{2} c_{\rm NFW}(M_{200}) + \frac{1-\alpha}{9}r_{200}\right), \delta \right],
 \label{eq:gNFWc} 
\end{equation}
where $c_{\rm NFW}(M_{200})=7.192\, (M_{200}/(10^{14}{\rm M}_\odot/h)^n$ with $n=0.114\pm 0.15$ \citep{DK15} and an arbitrary small $\delta >0$. We also consider the Einasto profile \citep{Ein} instead of the gNFW profile, as described in \citet{CBS19}.

\begin{deluxetable}{ccc}
  \tablecaption{Relation between stellar mass and halo mass used for our galaxy modeling in the $\Lambda$CDM framework, determined from weak-lensing measurements of \cite{Man16}. \label{tab:MsM200}}
\tablewidth{0pt}
\tablehead{
  \colhead{$\log_{10}(M_\star^{\rm Krou}/{\rm M}_\odot)$}  &  \colhead{$\log_{10}(M_{200}^{\rm WL}/{\rm M}_\odot)$} & \colhead{used uncertainty} \\
  \colhead{(1)}  &  \colhead{(2)}  &  \colhead{(3)}
}
\startdata
 $10.39$ &  $12.325^{+0.19}_{-0.24}$ & $\sqrt{0.22^2+0.083^2}$ \\
 $10.70$ &  $12.295^{+0.12}_{-0.14}$ & $\sqrt{0.13^2+0.083^2}$ \\
 $10.97$ &  $12.655^{+0.04}_{-0.05}$ & $\sqrt{0.05^2+0.083^2}$ \\
 $11.20$ &  $13.045^{+0.04}_{-0.04}$ & $\sqrt{0.04^2+0.083^2}$ \\
 $11.38$ &  $13.405^{+0.03}_{-0.03}$ & $\sqrt{0.03^2+0.083^2}$ \\
 $11.56$ &  $13.785^{+0.03}_{-0.03}$ & $\sqrt{0.03^2+0.083^2}$ \\
 $11.75$ &  $14.205^{+0.05}_{-0.05}$ & $\sqrt{0.05^2+0.083^2}$ \\
\enddata
\tablecomments{(1) MPA-JHU stellar mass for the Kroupa IMF. (2) Estimated weak-lensing mass within a sphere of $r_{200}$. (3) Actually used uncertainties: the additional factor of $0.083$ is included to account for the uncertainty associated with $M_\star^{\rm Krou}$. For $\log_{10}(M_\star^{\rm Krou}/M_\odot)\ga 11$, these values are a factor of 2 smaller than what were shown in Figure~1 of \citet{CBS18}. These choices of uncertainties do not affect our results.}
\end{deluxetable}

For MOND, we simply use the relation $a_{\rm X}(r)/a_{\rm B}(r) = f-1$ (Equation~\ref{eq:XM}), and we study a variety of choices for $a_0$ and the functional form of $f$ (Equations~(4-6)). The parameters in $\Lambda$CDM or MOND and their priors/constraints are summarized in Table~\ref{tab:prmt}.

\begin{deluxetable}{cccc}
\tablecaption{Parameters and their priors or constraints in $\Lambda$CDM or MOND. \label{tab:prmt}}
\tablewidth{0pt}
\tablehead{
  \colhead{parameter}  &  \colhead{free?}  &  \multicolumn{2}{c}{prior or constraint} \\
  \cline{3-4}
 \colhead{stars \& BH} &  \colhead{} & \colhead{$\Lambda$CDM} & \colhead{MOND} 
}
\startdata
 $\Upsilon_{\star 0}$ & free & \multicolumn{2}{c}{$>0$}  \\
 $K$ & free &  \multicolumn{2}{c}{$[0,~1.5]$}  \\
 $\beta_0$ & free &  \multicolumn{2}{c}{$[-2,~0.7]$}  \\
 $\beta_\infty$ & free &  \multicolumn{2}{c}{$[-2,~0.7]$}  \\
$r_a/R_{\rm e}$ & free &  \multicolumn{2}{c}{$[0.1,~1]$}  \\
$M_{\rm BH}$ & constrained &  \multicolumn{2}{c}{$M_{\rm BH}$-$\sigma_{\rm e}$ relation$^\ast$}  \\
  \cline{1-4}
 \colhead{DM halo} &  \colhead{} & \colhead{} & \colhead{} \\
  \cline{1-4}
$\alpha$  & free &  \colhead{$[0.1,~1]$} & \colhead{} \\
$M_{200}$  & constrained &  \colhead{Table~\ref{tab:MsM200}} & \colhead{} \\
$c_{200}$  & constrained & \colhead{Eq.~(\ref{eq:gNFWc})} & \colhead{} \\
  \cline{1-4}
 \colhead{MOND IF} &  \colhead{} & \colhead{} & \colhead{} \\
  \cline{1-4}
$\nu^\dagger$  & free &  \colhead{} & \colhead{$[0.1,~2]$} \\
(or $\lambda^\ddagger$)  &  &  \colhead{} & \colhead{(or $[0.3,~1.7]$)} \\
$a_0$ {\scriptsize [$10^{-10}$ m s$^{-2}$]}  & free &  \colhead{} & \colhead{$[0.5,~1.9]$} \\
\enddata
\tablecomments{$^\ast$See \citet{CBS18}. $^\dagger$MOND IF given by Equation~(\ref{eq:IFnu}). $^\ddagger$MOND IF given by Equation~(\ref{eq:IFlam}).}
\end{deluxetable}

We noted in the Introduction that the RAR carries qualitatively different implications for $\Lambda$CDM than for MOND.  Therefore, in what follows, we first determine the RAR by using the Jeans equation to explain the observed line-of-sight velocity dispersion in the $\Lambda$CDM paradigm.  We then follow the usual practice of asking what MOND IF best mimics the estimated dependence of $a_{\rm X}/a_{\rm B}$ on $a_{\rm B}$.  However, we then repeat the entire analysis under the MOND paradigm.  This second step is novel.  

\section{Results: The RAR in $\Lambda$CDM} \label{sec:lcdm}

In this section, we use a Bayesian approach (see, e.g., \citealt{WJ12}) to estimate the PDFs of the free parameters of a given model in $\Lambda$CDM. The free and constrained parameters are summarized in Table~\ref{tab:prmt}. The constrained parameters are, in principle, sampled in advance for each galaxy.  We will consider two methods of MC sampling as described below in Section~\ref{sec:bayes}. In one method, for an efficient sampling we will treat the constraint on $M_{200}$ (Table~\ref{tab:MsM200}) as a datum to be included in a likelihood function and then treat $M_{200}$ as a free parameter.

\subsection{Bayesian analysis of \atl\ \siglos} \label{sec:bayes}
Let $\vec{\Theta}$ denote the vector of free parameters associated with a model.  E.g., for the gNFW model, $\vec{\Theta} \equiv (\Upsilon_{\star 0}$, $K$, $\beta_0$, $\beta_{\infty}$, $r_a$, $\alpha$(, $M_{200}$)).  We then define 
\begin{eqnarray}
  \chi^2(\vec{\Theta}) & \equiv & \sum_{i=1}^{N_{\rm bin}} \frac{\left[\sigma_{\rm los}^{\rm obs}(R_i)-\sigma_{\rm los}^{\rm mod}(R_i;\vec{\Theta})\right]^2}{s_i^2} \nonumber \\ 
 & & \left( + \frac{[\log_{10} M_{200}- \log_{10} M_{200}^{\rm WL}]^2}{s_{\log_{10}M_{200}}^2} \right) ,
 \label{eq:chisq}
\end{eqnarray}
where $\sigma_{\rm los}^{\rm obs}(R_i)$ is the measured velocity dispersion with its uncertainty $s_i$ at $R_i$ and $\sigma_{\rm los}^{\rm mod}(R_i;\vec{\Theta})$ is the value predicted by the model under consideration when the model parameters are $\vec{\Theta}$. The last term (in round parenthesis) in Equation~(\ref{eq:chisq}) in only included if $M_{200}$ is treated as a free parameter. Note here that $M_{200}^{\rm WL}$ denotes the weak lensing constrained value given in Table~\ref{tab:MsM200}.

It is common to use this $\chi^2$ to define a likelihood. We consider two likelihoods. One is the commonly adopted Gaussian likelihood given by
\begin{equation}
  \mathcal{L}_{\rm Gaussian}(\vec{\Theta}) \propto \exp(-\chi^2(\vec{\Theta})/2).
 \label{eq:L}
\end{equation}
    In Bayesian inference, this likelihood, along with the prior assumptions about the range over which the parameters $\vec{\Theta}$ can be varied, is then used to generate MC samples (see, e.g., \citealt{AD14}) of the posterior distribution, which is the product of the prior and the likelihood.  We use a public Markov Chain Monte Carlo (MCMC) sampler `emcee', which is an implementation of the affine-invariant ensemble sampler by \citet{FM13}, to generate draws from this posterior distribution.  In principle, this provides the Bayesian inference of the posterior PDF of any of our free parameters or calculable quantities (e.g., $a_{\rm X}/a_{\rm B}$). Note that the posterior PDF returned by the emcee code, in general, differs from a Gaussian shape (i.e.\ Equation~(\ref{eq:L})) although we use flat priors for the free parameters (Table~\ref{tab:prmt}).

Although the likelihood given by Equation~(\ref{eq:L}) is widely used to infer posterior PDFs, there is no guarantee that an MCMC sampling based on it will produce an unbiased result for our RAR problem, i.e.\ constraining the ratio $a_{\rm X}/a_{\rm B}$ as a function of $a_{\rm B}$. In fact, our test of the MCMC sampling using mock velocity dispersion profiles indicates that the Gaussian likelihood can produce too narrow a posterior distribution and thus may be vulnerable to biases in some cases (see Appendix~\ref{sec:mc2mc} for details).

Therefore, we also consider a simple-minded `likelihood' given by
\begin{equation}
  \mathcal{L}_{\rm TopHat}(\vec{\Theta}) \propto
  \begin{cases}
 1 &\text{if} \hspace{1em} \bar{\chi}^2(\vec{\Theta})< \bar{\chi}^2_{\rm crit} \\
    0 & \text{else},
 \label{eq:Ltop}
  \end{cases}
\end{equation}
where $\bar{\chi}^2 \equiv \chi^2/N_{\rm dof}$ ($N_{\rm dof}=N_{\rm bin}-N_{\rm free}$) and $\bar{\chi}^2_{\rm crit}$ is a critical value that we define shortly.  Compared to Equation~(\ref{eq:L}), this does not penalize models (based on $\chi^2$) which provide worse fits:  they are either acceptable or not. We call this the Simple Monte Carlo (hereafter SMC) method. In this case, using a code written by one of us we search the entire parameter space robustly. Also, all constrained parameters including $M_{200}$ are sampled separately and thus the last term in Equation~(\ref{eq:chisq}) is not necessary. We note that the associated posterior PDF does not have a top-hat shape (even though the likelihood is effectively top-hat and the priors are uniform) because only certain combinations of parameters are accepted by Equation~(\ref{eq:Ltop}) and the PDF of $\bar{\chi}^2$ within these accepted models is not a top-hat shape. In general, the SMC method will lead to a broader posterior -- and hence to less restrictive constraints -- compared to the $\mathcal{L}_{\rm Gaussian}$-based MCMC method. In Section~\ref{sec:veri}, we verify that the SMC method produces robust RAR results in the MOND framework (see also Appendix~\ref{sec:mc2mc}).

\begin{figure}[t] %
 \centering
  \includegraphics[angle=0,origin=c,scale=0.45]{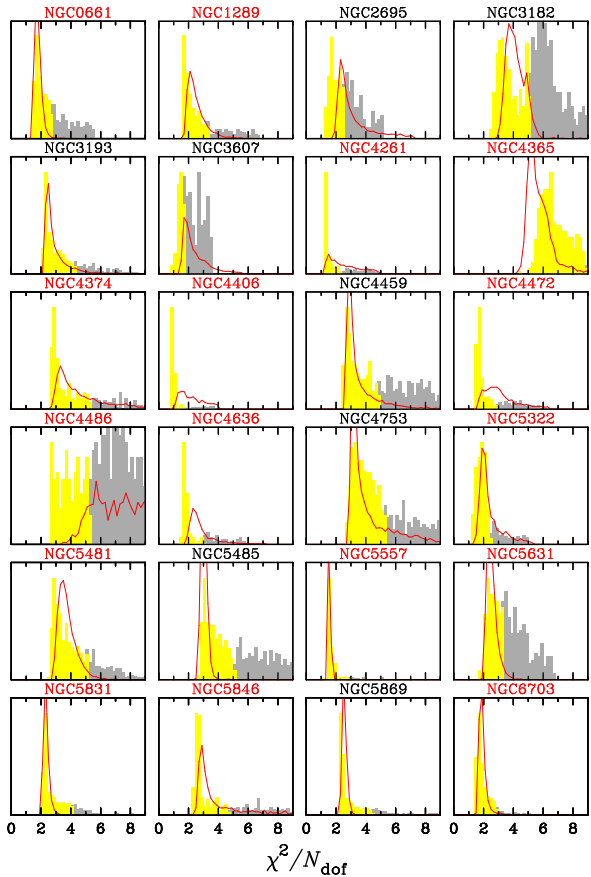}
  \caption{PDFs, from the SMC sampler, of the reduced $\chi^2$, i.e.\ $\chi^2$ (c.f.\ Equation~(\ref{eq:chisq})) per $N_{\rm dof}$ (degree of freedom) when fitting a model which has a gNFW DM halo (Equation~\ref{eq:gNFW}). Yellow regions show $\bar{\chi}^2< 2\bar{\chi}^2_{\rm min}$ while the gray regions are for $2\bar{\chi}^2_{\rm min}\le\bar{\chi}^2< 4\bar{\chi}^2_{\rm min}$.  Red curve in each panel shows the posterior $\bar\chi^2$ distribution obtained from the MCMC sampler. Each red curve has been normalized so that the area under it matches the area under the corresponding yellow histogram. In this and following figures slow rotators are named in red unless noted otherwise. \label{chisqred}}
\end{figure}

Figure~\ref{chisqred} shows the posterior distributions of $\bar{\chi}^2$ based on the two MC sampling methods. The red curve in each panel shows the distribution obtained from the MCMC sampler. The yellow histograms show models satisfying $\Delta \bar{\chi}^2 = \bar{\chi}^2 - \bar{\chi}^2_{\rm min}< \bar{\chi}^2_{\rm min}$ (where $\bar{\chi}^2_{\rm min}$ refers to the global minimum value from our SMC search), i.e., use $\bar{\chi}^2_{\rm crit}=2\bar{\chi}^2_{\rm min}$ in Equation~(\ref{eq:Ltop}). Because $\bar{\chi}^2_{\rm min} \lesssim 2.5$ for most galaxies, this means that the probability $P(\Delta \bar{\chi}^2>\bar{\chi}^2_{\rm min})$ would be $\ga 0.001$ for Gaussian statistics. This is a normally accepted criterion, and our numerical experiments, using mock velocity dispersion profiles, also suggest that this is a reasonable choice (see Section~\ref{sec:veri} and Appendix~\ref{sec:mc2mc}). However, we will also consider a relaxed criterion of $\bar{\chi}^2_{\rm crit}=4\bar{\chi}^2_{\rm min}$ (gray regions) to see the effect of varying $\bar{\chi}^2_{\rm crit}$. When we derive a statistically weighted RAR using all galaxies in Section~\ref{sec:stackRAR}, we will see that both choices of $\bar{\chi}^2_{\rm crit}$ give similar results (although individual RARs can be broader with $\bar{\chi}^2_{\rm crit}=4\bar{\chi}^2_{\rm min}$ in some galaxies).

\begin{figure*}
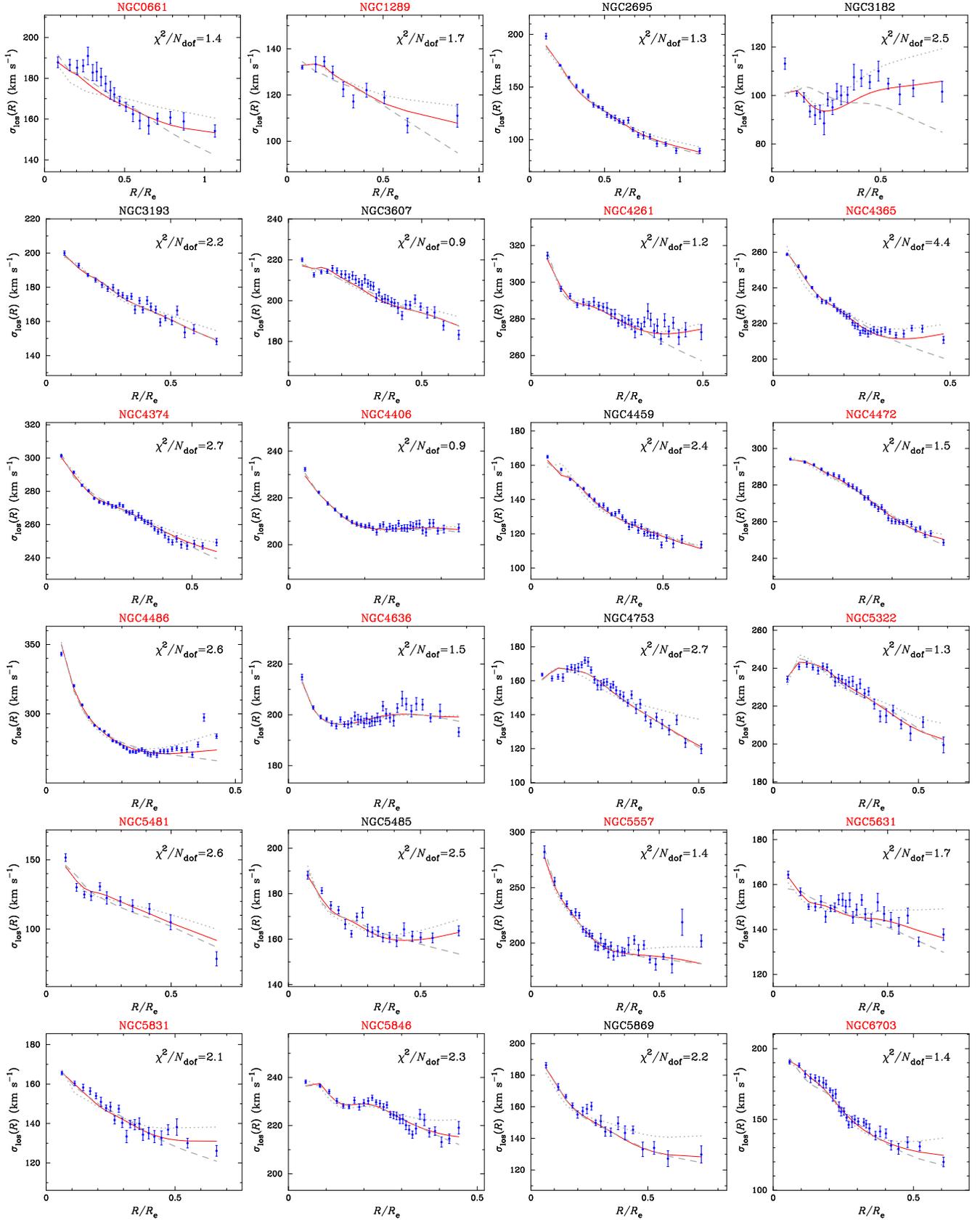
 %
  \hspace{-1ex}
  \includegraphics[angle=-90,origin=c,scale=0.207]{Fig_VP_NGC0661.eps}
  \includegraphics[angle=-90,origin=c,scale=0.207]{Fig_VP_NGC1289.eps}
  \includegraphics[angle=-90,origin=c,scale=0.207]{Fig_VP_NGC2695.eps}
  \includegraphics[angle=-90,origin=c,scale=0.207]{Fig_VP_NGC3182.eps}
  
  \vspace{-2ex}
  \includegraphics[angle=-90,origin=c,scale=0.207]{Fig_VP_NGC3193.eps}
  \includegraphics[angle=-90,origin=c,scale=0.207]{Fig_VP_NGC3607.eps}
  \includegraphics[angle=-90,origin=c,scale=0.207]{Fig_VP_NGC4261.eps}
  \includegraphics[angle=-90,origin=c,scale=0.207]{Fig_VP_NGC4365.eps}
  
  \vspace{-2ex}
  \includegraphics[angle=-90,origin=c,scale=0.207]{Fig_VP_NGC4374.eps}
  \includegraphics[angle=-90,origin=c,scale=0.207]{Fig_VP_NGC4406.eps}
  \includegraphics[angle=-90,origin=c,scale=0.207]{Fig_VP_NGC4459.eps}
  \includegraphics[angle=-90,origin=c,scale=0.207]{Fig_VP_NGC4472.eps}
  
  \vspace{-2ex}
  \includegraphics[angle=-90,origin=c,scale=0.207]{Fig_VP_NGC4486.eps}
  \includegraphics[angle=-90,origin=c,scale=0.207]{Fig_VP_NGC4636.eps}
  \includegraphics[angle=-90,origin=c,scale=0.207]{Fig_VP_NGC4753.eps}
  \includegraphics[angle=-90,origin=c,scale=0.207]{Fig_VP_NGC5322.eps}

  \vspace{-2ex}
  \includegraphics[angle=-90,origin=c,scale=0.207]{Fig_VP_NGC5481.eps}
  \includegraphics[angle=-90,origin=c,scale=0.207]{Fig_VP_NGC5485.eps}
  \includegraphics[angle=-90,origin=c,scale=0.207]{Fig_VP_NGC5557.eps}
  \includegraphics[angle=-90,origin=c,scale=0.207]{Fig_VP_NGC5631.eps}
  
  \vspace{-2ex}
  \includegraphics[angle=-90,origin=c,scale=0.207]{Fig_VP_NGC5831.eps}
  \includegraphics[angle=-90,origin=c,scale=0.207]{Fig_VP_NGC5846.eps}
  \includegraphics[angle=-90,origin=c,scale=0.207]{Fig_VP_NGC5869.eps}
  \includegraphics[angle=-90,origin=c,scale=0.207]{Fig_VP_NGC6703.eps}

  \vspace{-2em}
  \caption{Observed line-of-sight velocity dispersion profiles (symbols).  Smooth red curves show the best-fit models in which the DM halo is the gNFW of Equation~(\ref{eq:gNFW}).  Gray dotted and dashed curves show the models which give rise to the upper-most and lower-most RARs in Figure~\ref{RARgNFW} below.}
 \label{VDgNFW}
\end{figure*}  

The difference between the red curve and the shaded region in each panel of Figure~\ref{chisqred} illustrates how badly a model can fit the data and still be accepted by our two methods. Therefore, as a sanity check, Figure~\ref{VDgNFW} compares the observed \siglos\ profiles (symbols) with the gNFW models having $\chi^2_{\rm min}$.  Notice that the profiles which rise slightly at larger $R$ (i.e.\ NGC~3182, 4365, 4486) tend to be harder to fit (they have broader $\bar\chi^2$ distributions). The gray dotted and dashed curves show fits which give rise to the upper-most and lower-most RARs in Figure~\ref{RARgNFW} below.

\begin{figure} %
  \centering
  \includegraphics[angle=0,origin=c,scale=0.45]{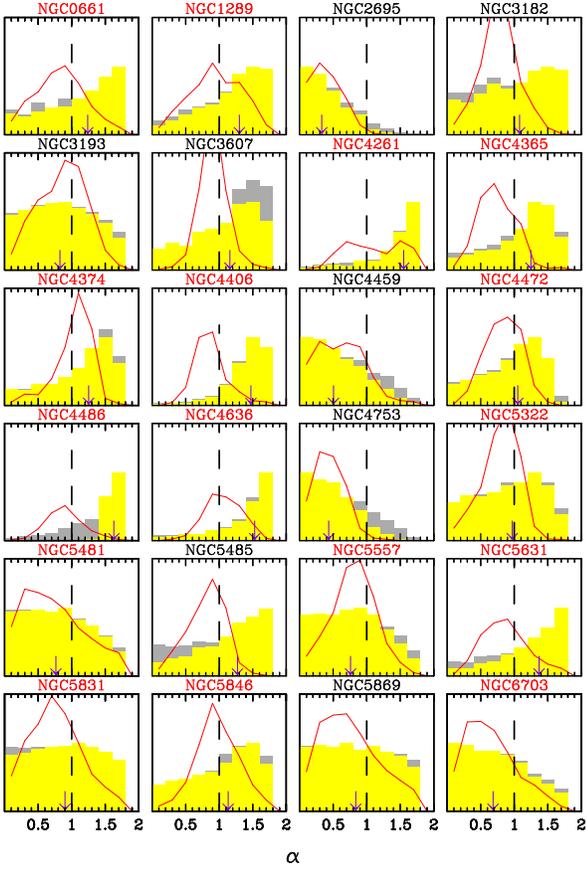}
  \caption{Posterior PDFs of the gNFW parameter $\alpha$ (Equation~(\ref{eq:gNFW})) profile from the SMC sampling of models which resulted in Figures~\ref{chisqred} and~\ref{VDgNFW}. Yellow regions represent models with $\bar{\chi}^2< 2\bar{\chi}^2_{\rm min}$ and gray regions represent $2\bar{\chi}^2_{\rm min}\le \bar{\chi}^2< 4\bar{\chi}^2_{\rm min}$. Black dashed vertical line corresponds to the NFW profile ($\alpha=1$) and the downward pointing arrows indicate the median values. Red curves show the corresponding distributions returned by the MCMC sampler.}
\label{alpha}
\end{figure}

\begin{figure} %
  \hspace{-1em}
  \includegraphics[angle=0,origin=c,scale=0.45]{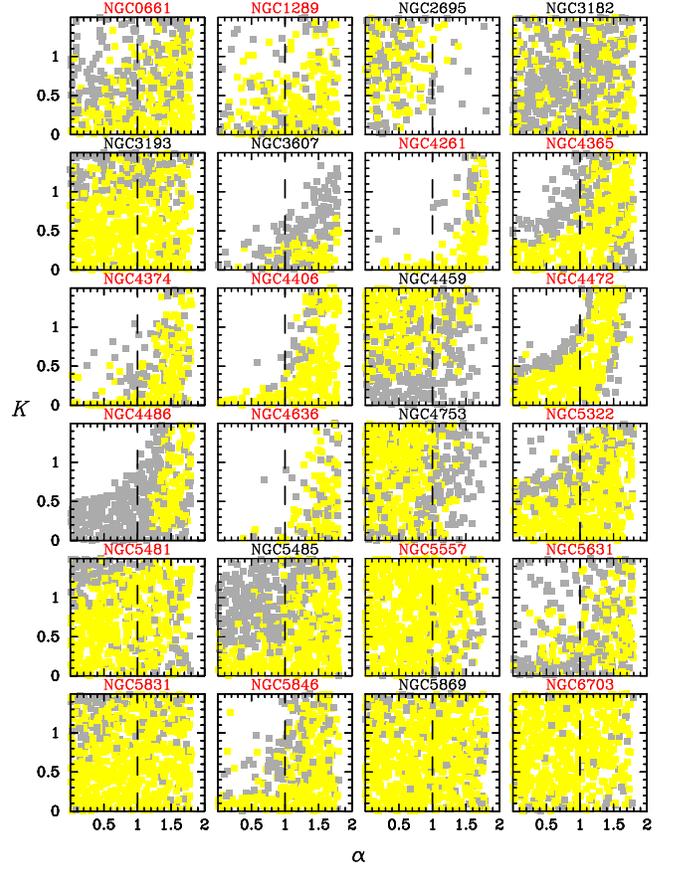}
  \caption{Correlation of $\alpha$ with $K$ from our SMC sampler: yellow squares represent models with $\bar{\chi}^2< 2\bar{\chi}^2_{\rm min}$ while the gray squares represent $2\bar{\chi}^2_{\rm min}\le \bar{\chi}^2< 4\bar{\chi}^2_{\rm min}$.  In many cases $\alpha$ increases when $K$ increases, but the opposite does not occur. Appendix~\ref{sec:corner} shows the full correlation of all parameters derived from the MCMC sampler for the four roundest galaxies.  \label{alphaK}}
\end{figure}

\subsection{Implications for halo profiles}\label{sec:alpha}
Since an MC set automatically defines a distribution of models (i.e.\ correlated distribution of free parameters), we can calculate the PDF of any quantity of interest.  Posterior distributions of stellar mass and $K$ based on our MC sets can be found in \citet{CBS18} while those of velocity dispersion anisotropy profiles can be found in \citet{CBS19}. In those papers we show that whenever an appropriate comparison is possible, our results agree well with independent literature results. 

Figure~\ref{alpha} shows the distribution of the gNFW-profile parameter $\alpha$ from our SMC sampling of the 24 \atl\ galaxies (gray histograms).  Red curves show the corresponding MCMC posteriors.  Figure~\ref{alphaK} shows that its value is correlated with that of the $\Upsilon_*$-gradient parameter $K$:  when $K$ increases, then $\alpha$ also tends to increase. The median value of $\alpha$ tends to be close to unity, suggesting that the simple NFW profile (as opposed to gNFW) is close to the median DM halo profile of elliptical galaxies.  This is consistent with recent results based on combined analyses of strong lensing and stellar kinematics \citep{Son15,Shan17}.

\begin{figure} %
  \vspace{-1em}
  \hspace{-5ex}
  \includegraphics[angle=0,origin=c,scale=0.50]{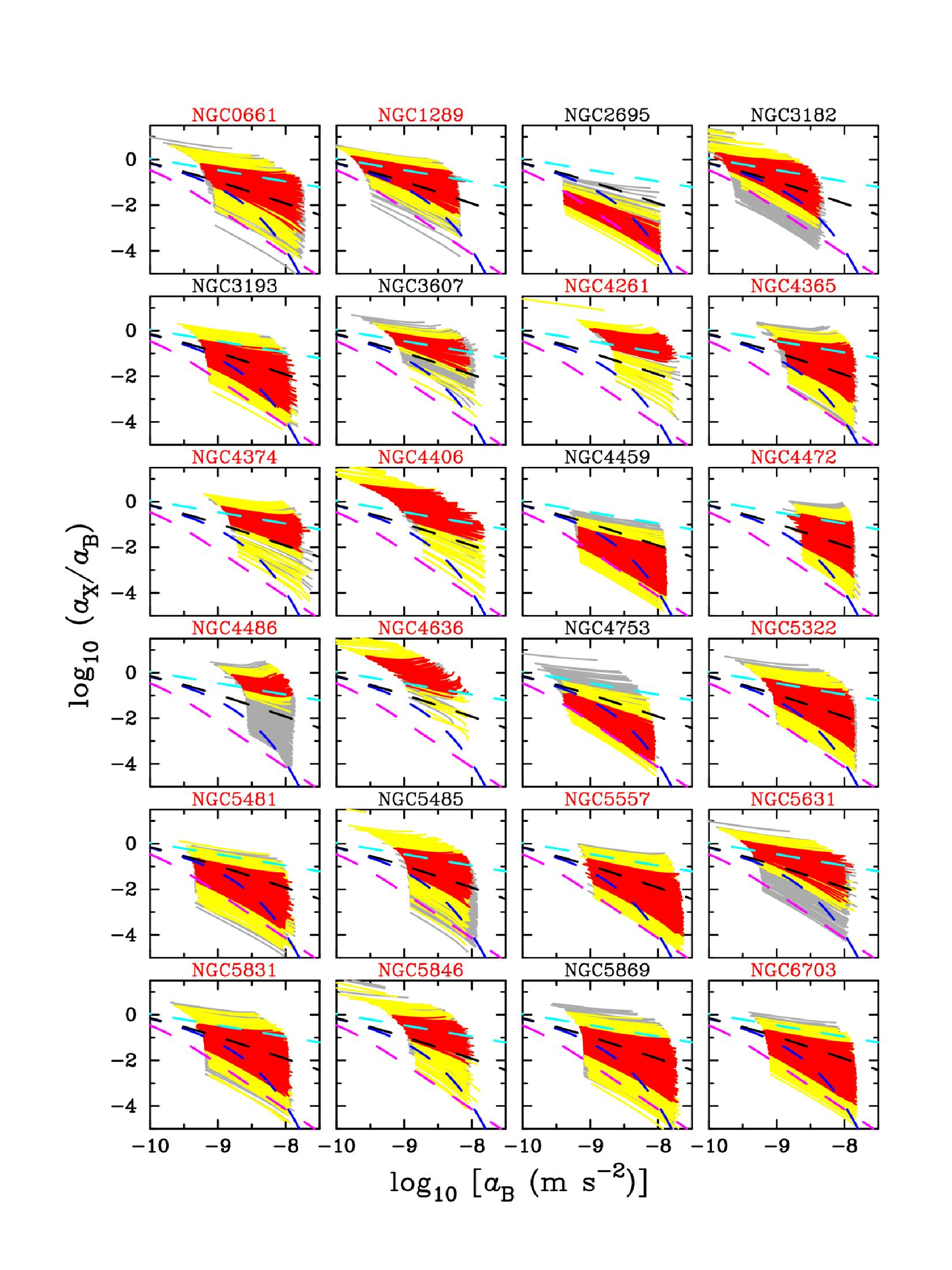}
  \vspace{-3em}
  \caption{Individual radial acceleration relation for 24 pure-bulge ATLAS$^{\rm 3D}$ galaxies when the DM component is assumed to follow the gNFW profile (Equation~(\ref{eq:gNFW})). Each colored line represents the radial behavior of the ratio $a_{\rm X}(r)/a_{\rm B}(r)$ from one model in the MC set. Yellow curves show cases that have $\bar{\chi}^2< 2\bar{\chi}^2_{\rm min}$; gray curves represent $2\bar{\chi}^2_{\rm min}\le\bar{\chi}^2< 4\bar{\chi}^2_{\rm min}$. Red curves represent the region which contains 68\% of the yellow lines. Thick colored dashed curves are the same functions shown in Fig.~\ref{IFs}. \label{RARgNFW}}
\end{figure}

\begin{figure}
  \vspace{-1em}
  \hspace{-5ex}
  \includegraphics[angle=0,origin=c,scale=0.5]{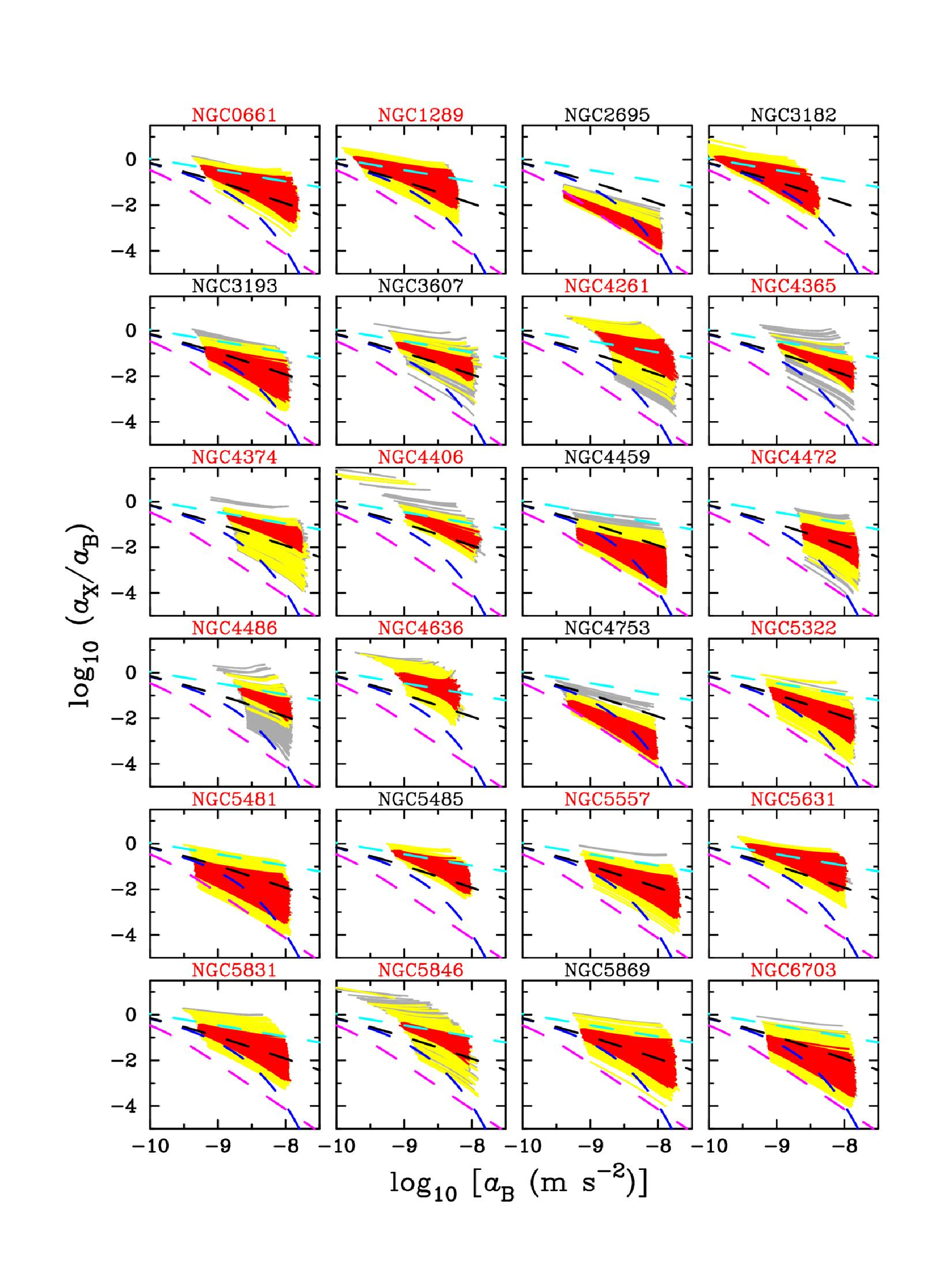}
  \vspace{-3em}
  \caption{Same as Figure~\ref{RARgNFW} but from the MCMC sampling.}
  \label{RARgNFWemcee}
\end{figure}

\subsection{Implications for the RAR}\label{sec:indRAR}

Let us directly consider our main question: what do our MC models imply for the RAR? This question in $\Lambda$CDM now appears to be particularly interesting because the median property of the constrained DM halo has been shown to be close to the NFW. Figure~\ref{RARgNFW} shows the distribution of the acceleration ratio $a_{\rm X}(r)/a_{\rm B}(r)$ from the SMC models for each of the 24 \atl\ galaxies.  The red curves are associated with model parameters which describe the observed \siglos\ best; yellow curves are slightly worse; and the gray curves have model parameters which provide poor descriptions of \siglos. Note that these individual results are more reliable for slow rotators (galaxy names in red) because we are using spherical galaxy models. Despite the wide range of \siglos\ shapes, the distribution of RAR shapes defined by the red curves is reasonably narrow.  This is the motivation for the suggestion that the RAR must encode interesting physics.  In the context of $\Lambda$CDM, this must constrain the physics of how these galaxies formed. 

Figure~\ref{RARgNFWemcee} shows the results from the MCMC sampling for the same model of the DM halo, i.e.\ the gNFW profile. Compared with the SMC results, the individual RARs are more narrowly distributed because the MCMC uses a Gaussian likelihood. If we use the Einasto profile for the DM halo (instead of the gNFW) whose parameterizations and priors are described in \citet{CBS19}, then we obtain essentially the same results. Clearly, the RAR is not sensitive to the differences between these two descriptions of DM halos.

\subsection{Implications for MOND}\label{sec:lcdm2mond}
 The RARs in $\Lambda$CDM shown in Figures~\ref{RARgNFW} and \ref{RARgNFWemcee} represent correlations (or the correlation) between baryons and the assumed DM under Newtonian gravity and Newtonian dynamics. However, if the DM halo approach is interpreted as simply a parameterization of the kinematic data, and the resulting RAR is not biased by the choice of parameterization, then we can actually use the RAR from the DM approach to learn something about other approaches. In this respect, it is particularly interesting that the RAR is directly related to the IF in the MOND paradigm. As the exact shape of the IF is a critical open question in MOND, it is interesting to compare the individual RARs from the DM approach to various proposed RARs or MOND IFs shown by thick colored dashed curves in Figures~\ref{RARgNFW} and \ref{RARgNFWemcee}.  This is the way in which a Newtonian+DM derived RAR is typically used.  However, we will also consider self-consistent analyses of MOND in Section~\ref{sec:mond}. 

Recall that, if MOND is correct, there is a single universal RAR which must provide an acceptable description of all galaxies.  That said, it appears that some galaxies (e.g.\ NGC 4406, 4486, 4636) disfavor the McGaugh IF while others (e.g.\ NGC 2695, 4459, 4753) disfavor the Bekenstein IF. On the other hand, it appears that the Simple IF does not have much difficulty in fitting any of the \atl\ galaxies except possibly for NGC~2695, which is however a fast rotator.  If a single IF must describe all galaxies, then the results of this section suggest that only the Simple IF is a viable model.  We discuss this further in Sections~\ref{sec:stackRAR} and Section~\ref{sec:chi2}). 

\subsection{The typical RAR from 4201 SDSS galaxies} \label{sec:RARsdss}
We now turn to the question of using the much larger sample of SDSS galaxies to estimate the RAR.  For these galaxies, the procedure for obtaining a MC set with the single dynamical constraint (i.e.\ $\sigma_{\rm ap}$: Equation~(\ref{eq:VDap})) necessarily differs from that for an \atl\ galaxy for which $\sigma_{\rm los}(R)$ is observed over a range of $R$.

For SDSS galaxies, the light distribution and DM halo parameters are drawn in the same way, but $K$ is drawn from a range $0\le K < K_{\rm max}$ where $K_{\rm max}$ is chosen so that the posterior distribution matches that for the \atl\ galaxies. Then, $M_{\rm \star e}$ is assigned using the fundamental mass plane (FMP) derived for the ATLAS$^{\rm 3D}$ galaxies, a correlation of the effective radius $R_{\rm e}$ with the light-weighted mean line-of-sight velocity dispersion $\sigma_{\rm e}\equiv \langle\sigma_{\rm los}\rangle(R=R_{\rm e})$ and the projected stellar mass $M_{\rm \star e}$  which is a function of $K$: $\log_{10} M_{\rm \star e}= \log_{10} M_{\rm \star e}(K) = \log_{10} M_{\rm \star e}(K=0)+b' K$ with $b'=-0.18^{+0.02}_{-0.07}$ \citep{CBS18}.  Finally, we search iteratively for a velocity dispersion anisotropy drawn randomly from a prior range (consistent with the \atl\ galaxies: see \citealt{CBS19}) predicting $\sigma_{\rm ap}$ within the estimated error.

In all, we generate a set of 90 MC models for each galaxy. If 68\% of the models in the set match the observed $\sigma_{\rm ap}$ within the formal measurement error, we accept the set. Only about 200 galaxies fail this test:  the vast majority, about 4000 galaxies, pass (see \citealt{CBS18} for further details).  For each of these models, we estimate $a_{\rm X}$ and $a_{\rm B}$ associated with the scale $R_{\rm ap}$.  From these, we generate posterior distributions of $a_{\rm X}$ and $a_{\rm B}$.

\begin{figure} %
  \includegraphics[angle=-90,origin=c,scale=0.33]{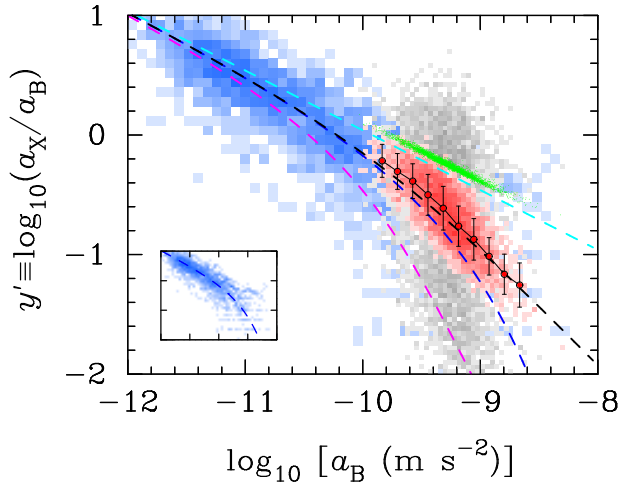}
  \vspace{-0.7cm}
  \caption{Gray pixels with varying intensity display the distribution of $a_{\rm X}/a_{\rm B}$ at $r=R_{\rm ap}$ as a function of $a_{\rm B}(r=R_{\rm ap})$ in one MC realization for 3991 SDSS galaxies for which modeling is successful with the standard input of $(K_{\rm max},b')=(1,-0.18)$ and a gNFW DM halo. Red pixels with varying intensity represent the median values in the 90 MC realizations of the 3991 galaxies. Red-filled circles and error bars represent the mean and standard deviation of the medians within each bin of $a_{\rm B}$. Colored dashed curves are the same as in Figure~\ref{IFs}. Green points are predicted by Verlinde's emergent gravity. Blue pixels represent 2693 data points from rotating galaxies in the literature \citep{Lel} for which we use the proxy $a/a_{\rm B}-1$ for $a_{\rm X}/a_{\rm B}$. The inset reproduces these data points only.}
\label{RARsdss}
\end{figure}

Figure~\ref{RARsdss} exhibits the distribution of $\sim 4000$ SDSS successfully modeled galaxies in the plane spanned by $a_{\rm B}$ and $a_{\rm X}/a_{\rm B}$ at $r=R_{\rm ap}$ of the galaxies for the standard input \citep{CBS18} under the $\Lambda$CDM case with the gNFW halo profile (if the Einasto profile is used instead, we obtain very similar results). In this figure we distribute MC models of the SDSS galaxies in pixels such that the intensity of each pixel indicates the occupancy of galaxies (i.e.\ a probability). Grey pixels represent one MC realization of the SDSS galaxies while red pixels represent the medians in the MC sets of the 90 realizations. The red pixels show that in agreement with the results for the \atl\ galaxies the Simple RAR (black dashed curve) is most consistent with SDSS galaxies. Figure~\ref{RARsdss} also reproduces published results for mostly spiral galaxies (blue pixels). Note that blue pixels occupy mostly sub-critical acceleration regions and are consistent with both the Simple and McGaugh's IFs but cannot distinguish them well in the supercritical regime.

\subsection{The Stacked RAR} \label{sec:stackRAR}

\begin{figure*}
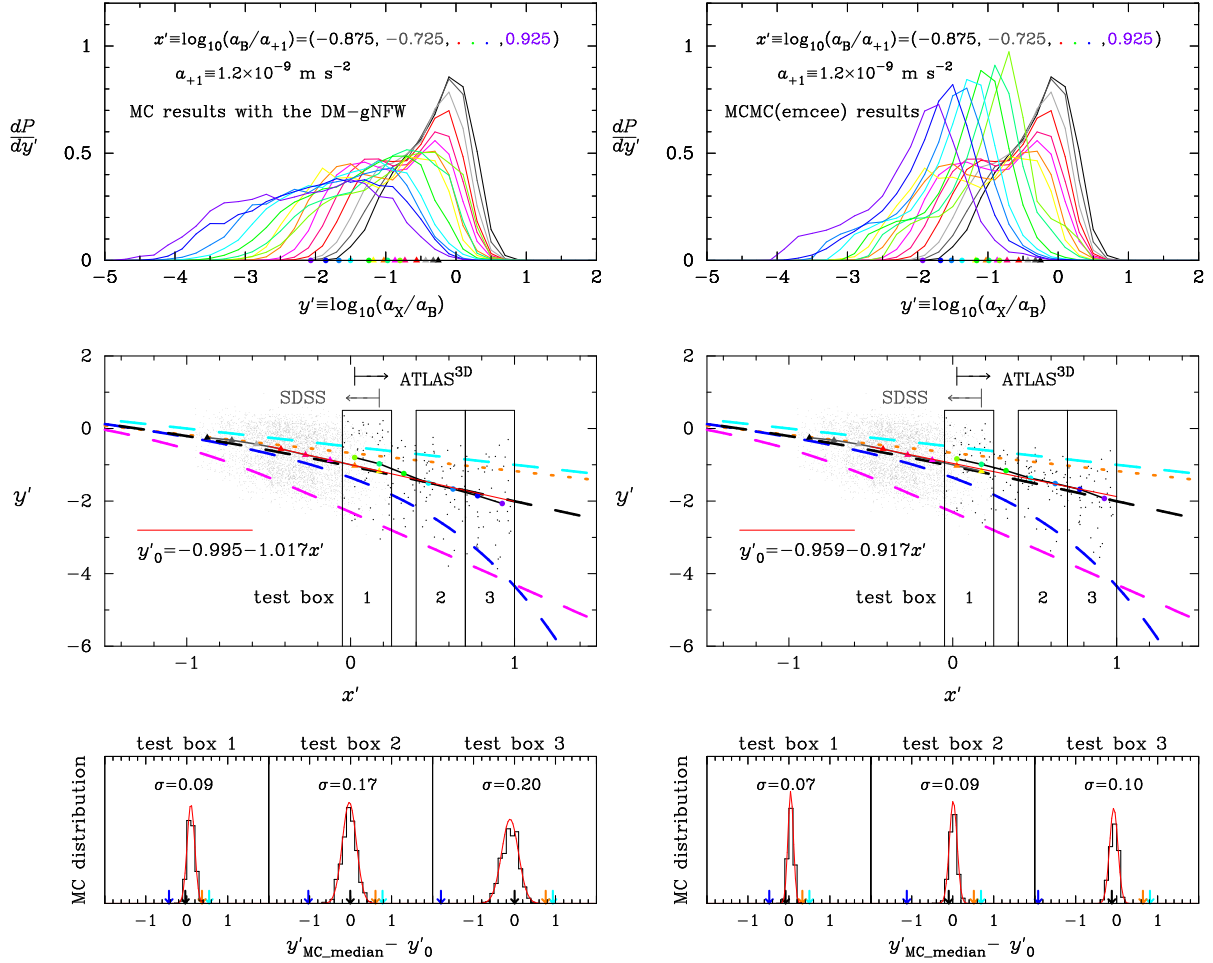

  \centering
  \includegraphics[angle=0,origin=c,scale=0.55]{Fig_PDFaXaB_gNFW.eps}
  \includegraphics[angle=0,origin=c,scale=0.55]{Fig_PDFaXaB_gNFW_emcee.eps}
  \caption{{\bf Left} - (Top) PDFs (Equation~(\ref{eq:PDF})) of $y'$ numerically constructed in the bins of $x'$ centered at $x'=-0.875$, $-0.725$, $\cdots$, $0.925$ based on the MC results with the gNFW DM halo profile (Equation~(\ref{eq:gNFW})) through the SMC sampling. Color-filled circles (triangles) show medians of the PDFs of the ATLAS$^{\rm 3D}$ (SDSS) galaxies. (Middle) - Stacked RAR (correlation of $y'$ with $x'$): Small dots show one MC realization for the considered bins of $x'$ based on the PDFs of the top panel. Colored circles and triangles show the medians in the bins of $x'$ displayed in the top panel.  Red line shows a linear fit to the points in the range $-0.5 <x' < 1$. Thick colored dashed curves show the same IFs as in Figure~\ref{IFs} with $a_0=1.2 \times 10^{-10}$~m~s$^{-2}$. The orange dotted curve represents the $\Lambda$CDM prediction from \citet{vP18}.
    (Bottom) - Distribution of the median of MC realizations which lie in the three test boxes labeled in the Middle panel. Red curve is a Gaussian fit to the histogram with the standard deviation indicated by $\sigma$. Downward pointing arrows indicate the predictions by the colored dashed/dotted curves at the relevant value of $x'$. {\bf Right} - Same as left, but through the MCMC sampling. }
\label{PDFdm}
\end{figure*}

So far we have obtained the distribution of MC models in the RAR plane for each of the 24 \atl\ galaxies and $\sim 4000$ SDSS galaxies (Figures~\ref{RARgNFW} ,\ref{RARgNFWemcee}, and \ref{RARsdss}). Figure~\ref{PDFdm} exhibits the outcome of statistically weighting all the individual results. We have obtained this as follows. First, because the typical acceleration scale in our sample is super-critical, we define 
$a_{+1}\equiv 10^{+1} \times (1.2 \times 10^{-10})$~m~s$^{-2}$ $=1.2 \times 10^{-9}$~m~s$^{-2}$, and express all accelerations in units of $a_{+1}$. We define 13 bins of $x' \equiv \log_{10}(a_{\rm B}/a_{+1})$ in steps of $0.15$ in the range $-1< x' < 1$.

We then calculate a statistically-weighted PDF of $y'\equiv \log_{10}(a_{\rm X}/a_{\rm B})$ in each bin of $x'$ using the SDSS (or \atl) galaxies whose individual RAR results belong to (or has an overlap with) the $x'$ bin. Note that an \atl galaxy spans multiple bins while an SDSS galaxy belongs to one bin. Note also that \atl\ galaxies occupy bins of higher $x'$ while SDSS galaxies occupy lower $x'$ (two middle bins are overlapped). For the $i$-th SDSS (or \atl) galaxy of the $x'$ bin we derive $dP_i(<y')/dy'|_{x'}$ using the models in the MC set (i.e.\ 90 MC realizations for each SDSS galaxy like the gray pixels as in Figure~\ref{RARsdss}, or all curves for each \atl\ galaxy as in Figure~\ref{RARgNFW}) which have $x'$ in the desired range. Here $P_i(<y')|_{x'}$ is the cumulative probability up to $y'$ at $x'$ (in the sense of $P_i(<\infty)=1$) calculated numerically using the MC models of the galaxy. For the $x'$ bin we derive a statistically-weighted (``stacked'') PDF
\begin{equation}
\left. \frac{dP(<y')}{dy'}\right|_{x'}\propto \sum_i \left. \frac{dP_i(<y')}{dy'}\right|_{x'} w_i ,
  \label{eq:PDF}
\end{equation}
where $P_i(<y')|_{x'}$ is the individual PDF of $i$-th galaxy and $w_i$ is the statistical weight assigned to that PDF. We have tried $w_i=1$ (i.e.\ uniform weighting) and $w_i=\exp\left(-\bar{\chi}_{{\rm min},i}^2/2\right)$, where $\bar{\chi}_{{\rm min},i}^2$ is the minimum value of the reduced $\chi^2$ for the galaxy. Although this second weighting scheme accounts for the fact that some MC realizations provide much better descriptions of the measured \siglos\ than others, it turns out that both weighting factors give very similar results.

The top panels of Figure~\ref{PDFdm} show the stacked PDFs of $y'$ for a number of bins in $x'$ (as labeled) when $w_i=\exp\left(-\bar{\chi}_{{\rm min},i}^2/2\right)$. The triangles and circles along the $x$-axis of these top panels show the medians of these PDFs based on SDSS and \atl\ galaxies, respectively.  The middle panels show the result of plotting these median values versus $x'$, and constitute our estimate of the `stacked' RAR associated with gNFW DM halo profiles based on the SMC sampling (left) or the MCMC sampling (right) in the $\Lambda$CDM paradigm.  Over the acceleration scales where they overlap, there is reasonable agreement between our SDSS and \atl\ samples.  Moreover, the estimated RAR does not depend strongly on whether we parameterize the DM distribution using a gNFW or Einasto profile.  

The dots in the middle panel show $x'$ and $y'$ values for one MC realization of each of the \atl\ and SDSS galaxies: i.e., each \atl\ galaxy contributes 7 (number of the $x'$ bins) dots, whereas each SDSS galaxy contributes only one. The filled triangles and circles show the median values obtained from the top panels for the two samples.

\subsection{Implications for $\Lambda$CDM and MOND from the stacked RAR} 

 Having estimated the empirical RAR from a Newtonian analysis, we now compare it with various predictions/suggestions including the $\Lambda$CDM prediction and MOND-related predictions. The philosophy here is that the RAR, if derived robustly from the kinematic data through any approach, may provide a useful constraint on any other approach. We do this in two ways.  First, we simply overplot the same four MOND IFs, using thick dashed curves, as in previous figures.  Recall that differences between the curves cannot be attributed to reasonable changes in $a_0$ (c.f.\ Figure~\ref{IFs}). The thick dotted orange curve shows a prediction for $\Lambda$CDM taken from \citet{vP18}; it is quite similar to Bekenstein's IF (cyan). It is clear that the Simple and McGaugh IFs (black and blue curves) and the $\Lambda$CDM prediction (orange dotted curve) agree below $10^{0.5}a_{+1}$ and provide good descriptions of our empirically determined RAR up to $\sim a_0$ probed by our SDSS galaxies. However, in the acceleration range which is dominated by \atl galaxies, the Simple IF provides a better description than the others.

Our second way of demonstrating this is as follows.  For the probed range $-1 < x' < 1$ the points defining the RAR cannot be described by a linear relation; some curvature is evident in the middle panels of Figure~\ref{PDFdm}. However, for the range $-0.5 < x' < 1$ we may approximate the RAR by a linear fit of $y'=p+q x'$ with $p = -1.00 \pm 0.03$ and $q=-1.02 \pm 0.09$ (the SMC case), or $p = -0.96 \pm 0.02$ and $q=-0.92 \pm 0.06$ (the MCMC case), where the statistical uncertainties have been estimated using a number of Monte Carlo realizations using the PDFs of the top panels. Although we do not show so explicitly, the result with Einasto DM halo profiles is very similar with $p = -1.07 \pm 0.03$ and $q=-0.99 \pm 0.10$  (Einasto, the SMC case).

The bottom panels show examples of the Monte Carlo distributions for three narrow ranges of $x'$. One of them (test box 1) uses 4 PDFs (two from the \atl\ galaxies and the other two from the SDSS galaxies) within the range of $x'$ while each of the other two (test box 2 and 3) uses 2 PDFs from the \atl\ galaxies within the respective range of $x'$. Downward pointing arrows indicate the predictions of the IFs represented by the colored dashed curves at the relevant value of $x'$. Both the linear fit for $-0.5 < x' < 1$ and the three test boxes clearly prefer the Simple RAR indicated by the black dashed curve in the middle panel and black downward pointing arrows in the test boxes. As the RAR determined from the data used Newtonian physics, the next section provides a fully self-consistent check of the conclusion that the Simple RAR provides the best fit of the MONDian IFs we have considered.
   
 However, it is fair to explore the discrepancy between the RAR and $\Lambda$CDM model represented by the orange dotted line.  The orange arrows in the bottom panels (see test boxes 2 and 3) of Figure~\ref{PDFdm} suggest that it is not consistent with our data.  This may seem contradictory, as our analysis was done within the $\Lambda$CDM framework.  The point is that this $\Lambda$CDM `prediction' depends on the assumed astrophysics of galaxy formation (represented, in this case, by the MUGS simulations analyzed by \citealt{vP18}).  The mismatch here suggests that the RAR provides an interesting complementary test of the physics of galaxy formation.
   
Finally, we point out that Figure~\ref{RARsdss} shows a comparison of the gNFW-based RAR we have determined for ellipticals with that for spirals. While the spiral galaxies data can be well described by either the McGaugh IF or the Simple IF, the elliptical galaxies can only be described by the Simple IF. This highlights the gain that comes from studying the super-critical regime with elliptical galaxies.

\section{Results:  The RAR in MOND}\label{sec:mond}
The previous section presented a Bayesian analysis of \siglos\ in elliptical galaxies which assumed Newtonian accelerations and parametric models for the dark matter profile.  Comparison of the resulting RAR with several MOND IFs suggested that the Simple IF (Equation~\ref{eq:IFnu} with $\nu=1$) provides a good description.  Here, we provide a complementary analysis of the same data which is performed entirely within the MOND framework.  Namely, instead of the parameters associated with the DM halo ($\alpha$, etc.) we work with the parameters of the IF (Equations~(\ref{eq:IFnu}) or~(\ref{eq:IFlam})):  $a_0$ and $\nu$ or $\lambda$.  

\subsection{Verification of methodology using mock profiles}\label{sec:veri}

In principle, the `free parameters' in MOND are not analogous to those of the DM, because a given $(a_0,\nu)$ or $(a_0,\lambda)$ pair must work for all galaxies.  (In contrast, there is no requirement that $\alpha$ be the same for all galaxies in the gNFW DM halo.)  Therefore, although it is straightforward to run our Bayesian analysis pipeline with $a_0$ and $\lambda$ or $\nu$ left as free parameters, it is not obvious what the results will mean.

To address this, we use the observed light profiles and the estimated $K$ and velocity dispersion anisotropy parameters (from modeling real \siglos\ profiles). We then set $(\lambda,a_0) = (1,1.2\times 10^{-10}~{\rm m~s}^{-2})$ to generate velocity dispersions, which we project and then add noise to get a mock \siglos.  By choosing these values of $(\lambda,a_0)$ we have ensured that our set of mock profiles represent plausible observations if the McGaugh IF was correct for all galaxies.  We then run the Bayesian analysis pipeline, leaving $(\lambda,a_0)$ as well as $\Upsilon_{\star 0}$, $K$, and anisotropy parameters as free, treating this mock \siglos\ as the data.  If the analysis pipeline is unbiased, it should recover the input $(\lambda,a_0)$.

\begin{figure} %
  \vspace{-1em}
  \hspace{-5ex}
   \includegraphics[angle=0,origin=c,scale=0.5]{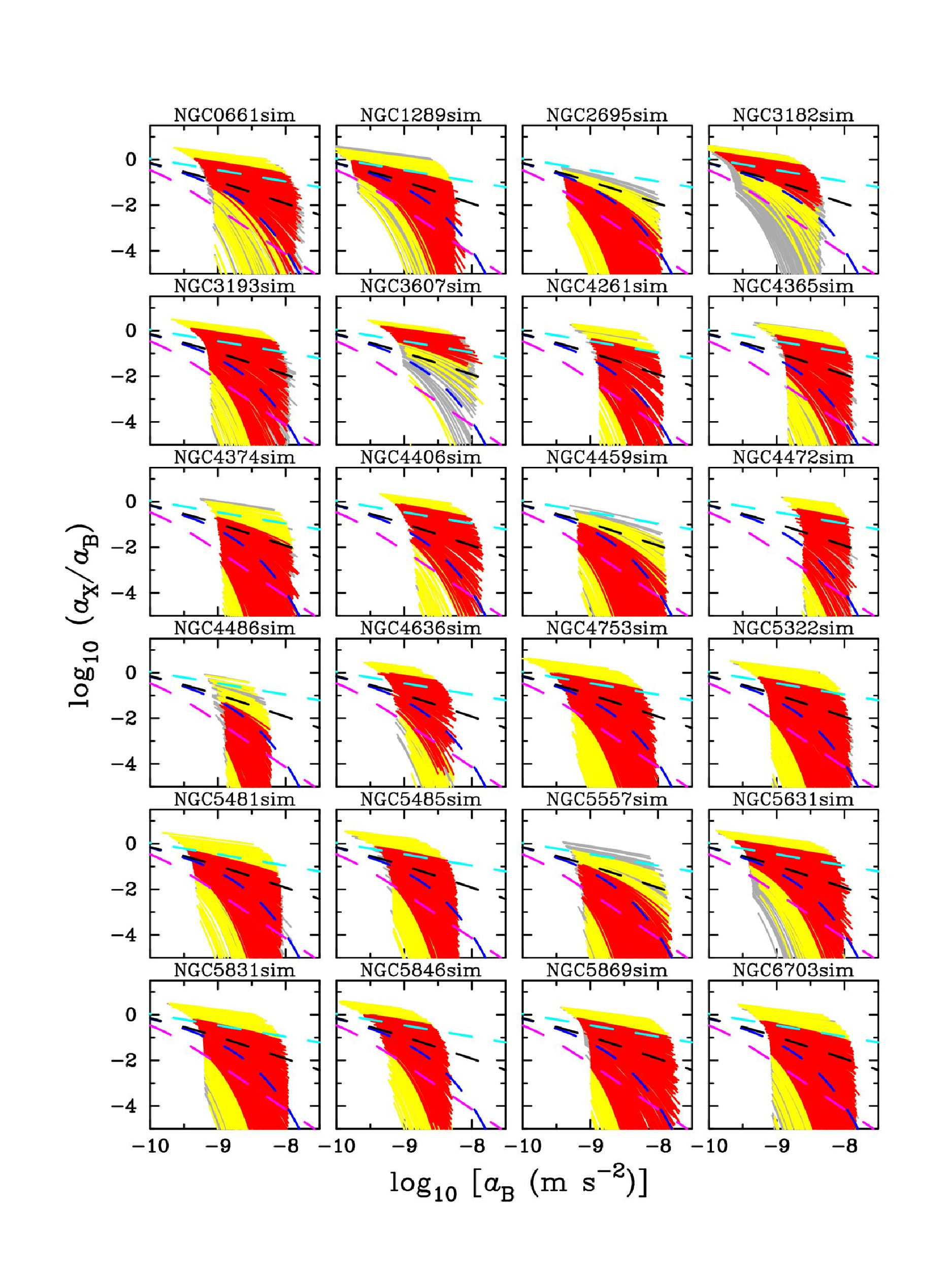}
  \vspace{-3em}
   \caption{Individual RARs derived from the mock McGaugh IF (Equation~(\ref{eq:IFlam}) with $\lambda=1$) \siglos\ profiles by our SMC analysis in which $\lambda$ and $a_0$ are free parameters.}
\label{RARsim}
\end{figure}

\begin{figure} %
  \centering
  \includegraphics[angle=0,origin=c,scale=0.55]{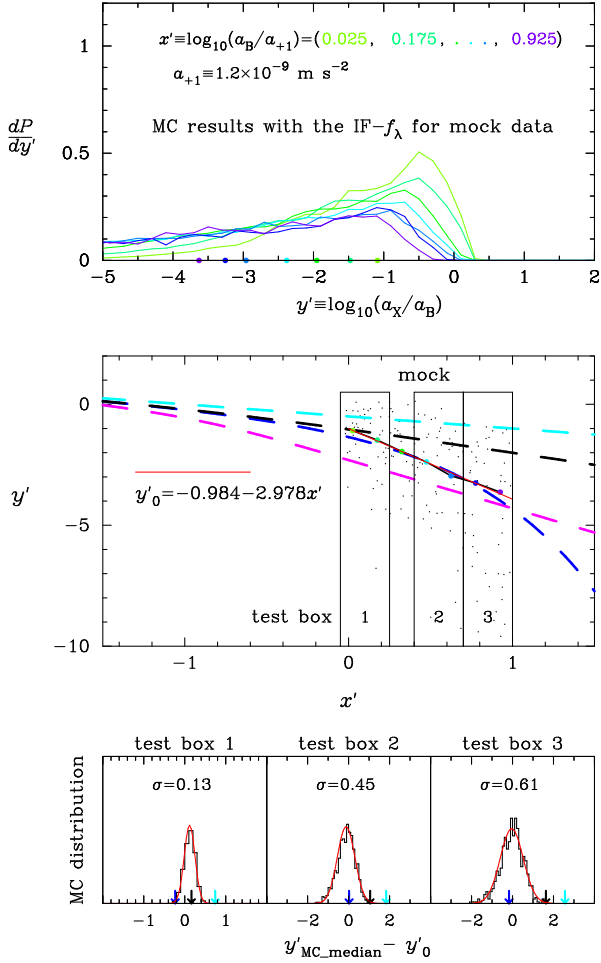}
  \caption{The stacked RAR determined from our analysis of the mock \siglos, shown in the same format as Figure~\ref{PDFdm}. SDSS data are not included here because we are considering mock data only.}
\label{PDFsim}
\end{figure}

Figures~\ref{RARsim} and~\ref{PDFsim} show the results through the SMC sampling.  Individual RARs (distributions of the SMC models in the $a_{\rm B}$-$a_{\rm X}/a_{\rm B}$ plane) are statistically consistent with the input McGaugh IF. That is to say, for all 24 cases the McGaugh IF is within the yellow regions defined by our acceptance criterion $\bar{\chi}^2 < 2 \bar{\chi}^2_{\rm min}$, and for over two thirds of cases the McGaugh IF is within the 68\% confidence region (red). Also, the stacked RAR based on the stacked PDFs of $y'$ as a function of $x'$ clearly recovers the McGaugh IF without any bias. In particular, the middle panel of Figure~\ref{PDFsim} shows that the stacked RAR is centered on the McGaugh IF, and the bottom panel `rules out' the Simple IF.  This suggests that if there is a universal RAR, and if it has a functional form which is similar to the MOND IFs, then our Bayesian analysis, in which we allow the parameters of the IF to vary from one object to another, will return an RAR which is close to the true universal one.  

 We have also obtained MCMC results using the code emcee for the mock velocity dispersions (see Appendix~\ref{sec:mc2mc}).  The MCMC results provide narrower (more precise) regions in the RAR space compared to SMC, but in some cases, the ``precise regions'' exclude the McGaugh IF, even though it is the correct answer. This means that the emcee code sometimes produces unrealistically narrow constraints. With this caveat in mind, we will present mainly the results from the SMC sampling for the real data in the MOND framework. However, it turns out that both sampling methods give consistent results with the rule of thumb that the MCMC sampling gives narrower results. 

\begin{figure}
  \centering
  \includegraphics[angle=0,origin=c,scale=0.55]{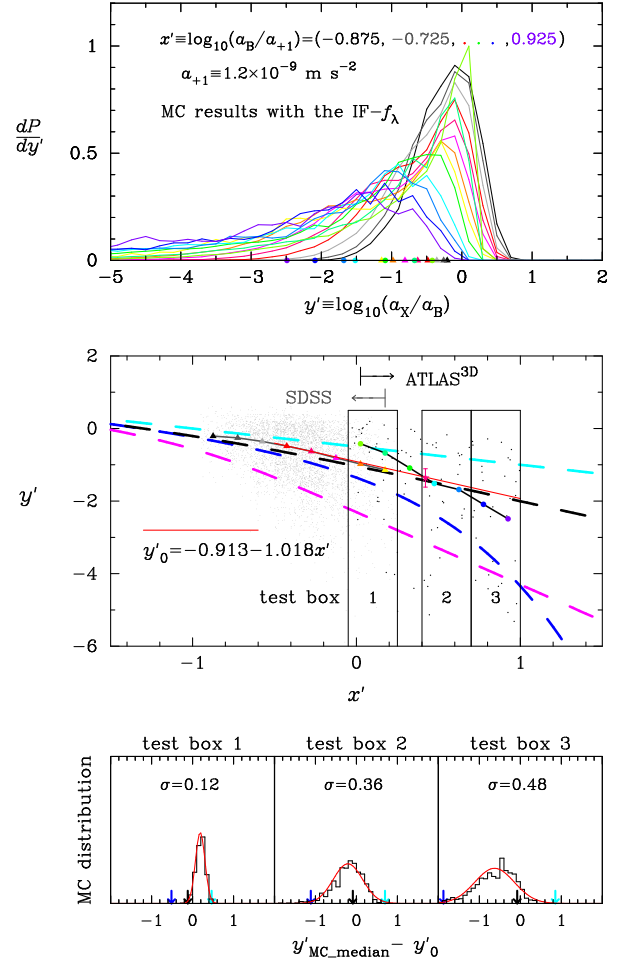}
\caption{Same as Figure~\ref{PDFsim} but for the \atl\ \siglos:  i.e., stacked RAR assuming the MOND IF is given by Equation~(\ref{eq:IFlam}) with the parameters $\lambda$ and $a_0$ allowed to vary from one galaxy to another.  The pink star with an error bar is the mean of the 24 best-fit models at $x'=0.42$ based on $f_\gamma$ (Equation~(\ref{eq:IFgam})), which provides a robust test of the Bekenstein (cyan) IF (see the text).
}
\label{PDFmondlam}
\end{figure}

\begin{figure}
  \centering
  \includegraphics[angle=0,origin=c,scale=0.55]{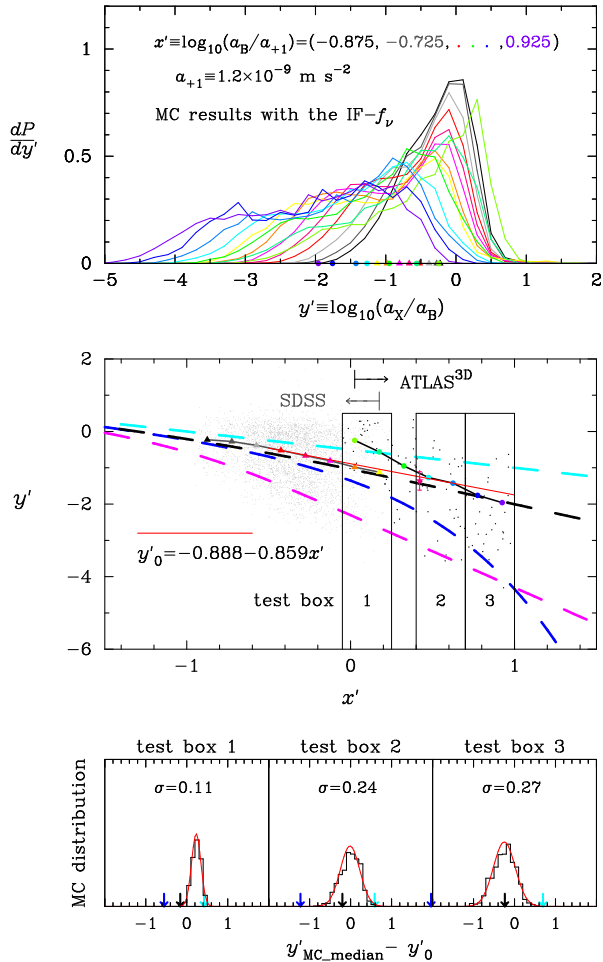}
\caption{Same as Figure~\ref{PDFmondlam} but with the MOND IF given by Equation~(\ref{eq:IFnu}) with parameters $\nu$ and $a_0$ allowed to vary from one galaxy to another.}
\label{PDFmondnu}
\end{figure}

\subsection{The stacked RAR}
With this in mind, we run our Bayesian SMC analysis pipeline on the \atl\ \siglos\ profiles, allowing both the MOND IF profile index $\lambda$ and the critical acceleration $a_0$ to be free parameters within specified prior ranges:  
$0.3 <\lambda < 1.7$ and $0.5<(a_0/10^{10}~{\rm m~s}^{-2})<1.9$.
We then obtain the ratio $a_{\rm X}(r)/a_{\rm B}(r)$ using the MC models of each galaxy, and combine them to produce a stacked RAR.  Figure~\ref{PDFmondlam} exhibits the results.  The middle and bottom panels show that McGaugh's IF is not picked out by the stacked RAR: the Simple IF is clearly favored.

Since the Simple IF is a special case of Equation~(\ref{eq:IFnu}) rather than~(\ref{eq:IFlam}), Figure~\ref{PDFmondnu} shows the result of repeating the entire analysis, but using Equation~(\ref{eq:IFnu}) as the IF.  In this case, we allow the same prior range in $a_0$, but allow $0<\nu \le 2$.  Not only is the Simple IF again preferred, but the distributions shown in the bottom panel are narrower than they were in Figure~\ref{PDFmondlam}, and are more like those in Figure~\ref{PDFdm}.  This again suggests that the Simple IF is closer to the true RAR than is McGaugh's function.

For completeness, a linear fit of $y'=p+q x'$ over the range $-0.5 < x' < 1$ returns
 $p = -0.91 \pm 0.04$ and $q=-1.02 \pm 0.13$
in the $f_\lambda$ case, and 
 $p = -0.89 \pm 0.04$ and $q=-0.86 \pm 0.12$ in the $f_\nu$ case.
These values are rather similar to those under the $\Lambda$CDM paradigm (Figure~\ref{PDFdm}).

We have also considered using the model $f_\gamma$ (Equation~(\ref{eq:IFgam})) because it can allow a more direct test of Bekenstein's IF. The pink star with an error bar displayed in Figures~\ref{PDFmondlam} and \ref{PDFmondnu} is the mean of the best-fit models of the 24 \atl\ galaxies with $f_\gamma$. As our modeling of mock velocity dispersion profiles generated with Bekenstein's IF verifies that this quantity is robustly reproduced, this result also prefers the Simple IF than other IFs.

To sum up, in the super-critical acceleration regime probed by the nearly spherical \atl\ and SDSS galaxies, the Simple IF is consistent with our stacked RAR, whereas Bekenstein's and McGaugh's IFs are not.  For each of test boxes 2 and 3 in the bottom panels of Figures~\ref{PDFmondlam} and~\ref{PDFmondnu}, the two IFs are excluded at $\sim 3\sigma$ to $\sim 9\sigma$ depending on the DM halo or MOND IF parameterization. This means that they are excluded at $\ga 4\sigma$.

\subsection{Universality of the MOND parameters}\label{sec:univ}
Although the Simple IF appears to provide a good description of the RAR, because $\nu$ and $a_0$ are supposed to be the same for all galaxies in the MOND paradigm, it is important to check whether the PDFs of $\nu$ and $a_0$ returned by our Bayesian analysis can support universality. 

The issue of universality in rotating galaxies was recently discussed in the context of whether $a_0$ is universal for all galaxies for a fixed MOND functional form \citep{Rod18,McG18,Kro18}. For such an analysis the algorithm employed to infer the posterior PDF plays a critical role because the conclusion can depend on the width of the PDF. Here we directly tackle the MOND functional form $f_{\nu}$ for a narrow range of $a_0=(1.0,~1.4) \times 10^{-10}$ m~s$^{-2}$. We intend to check whether $\nu=1$ is indeed consistent with all galaxies.

\begin{figure} %
  \centering
  \includegraphics[angle=0,origin=c,scale=0.45]{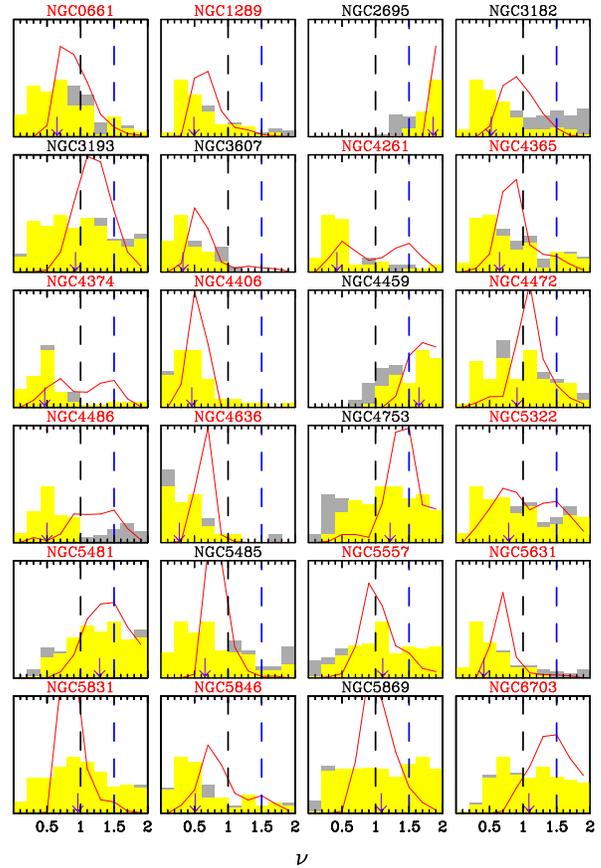}
  \caption{Posterior PDFs of $\nu$ for the $f_\nu$-IF (Equation~\ref{eq:IFnu}) from the models satisfying $1.0 < a_0 <1.4$ (in units of $10^{-10}$~m~s$^{-2}$).  Yellow (or gray) histograms represent the PDFs returned by our SMC sampling with $\bar{\chi}^2 < 2 \bar{\chi}^2_{\rm min}$ (or $\bar{\chi}^2 < 4 \bar{\chi}^2_{\rm min}$). Downward pointing arrows indicate the medians of the yellow histograms. Red curves show the PDFs returned by the MCMC sampling. Black dashed vertical line corresponds to the Simple IF while the blue line corresponds to the closest match to the McGaugh IF in this parameterization. \label{nu}}
\end{figure}

 Figure~\ref{nu} exhibits the individual PDFs of $\nu$ through the SMC or MCMC sampling. Overall, the two samplers give qualitatively consistent results, but the MCMC PDFs are narrower, as expected. (We have already noted, in Section~\ref{sec:veri} that the MCMC results can sometimes be unrealistically narrow.) The individual PDFs of $\nu$ from the SMC typically include the Simple IF ($\nu=1$) but exclude the McGaugh IF ($\nu=1.5$) in several cases. If we limit our attention to the more reliable sample of 16 slow rotators, our results are clearly consistent with the view that the Simple IF is universal in the sample. Given this preliminary result, it will be interesting to see future results from analyses of much larger samples. 

\begin{figure} %

 \vspace{1em}
  \includegraphics[angle=-90,origin=c,scale=0.4]{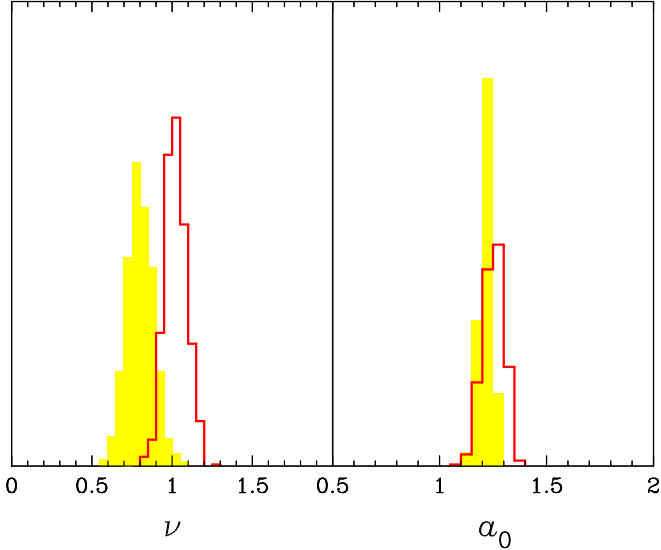}

 \vspace{-2em}
 \caption{ Monte Carlo distribution of ``universal'' values of $\nu$ (of the $f_{\nu}$ IF) and $a_0$ (in units of $10^{-10}$~m~s$^{-2}$) for the 24 \atl\ galaxies based on a bootstrap resampling of individual medians shown in Figure~\ref{nu}.  Yellow and red histograms show the SMC and MCMC results, respectively. Both are consistent with the Simple IF ($\nu=1$) with $a_0\approx 1.2$. \label{nua0}}
\end{figure}

Accepting the view that there exists a universal IF in our sample, we now determine the ``universal'' values of $\nu$ and $a_0$ of the assumed $f_\nu$ function in our sample. In doing so, we treat each galaxy as one datum, and the median of the PDF (the downward pointing arrow in each panel of Figure~\ref{nu}) represents a measured value of $\nu$ (and the width of the PDF represents the individual uncertainty). To determine the universal value and its uncertainty we use a bootstrap resampling of the individual medians shown in Figure~\ref{nu}. We obtain the sample mean of each bootstrap sample and the distribution of the sample means provides our estimate of the universal value. Figure~\ref{nua0} shows the results for $\nu$ and $a_0$. We obtain $\nu = 0.80\pm 0.08$ and $a_0=1.24\pm 0.04$ (based on the SMC results), or $\nu = 1.02\pm 0.07$ and $a_0=1.25\pm 0.05$~$10^{-10}$~m~s$^{-2}$ (based on the MCMC results). These results are in good agreement ($<2.4\sigma$) with the Simple IF ($\nu=1$) as expected. The case of $\nu=1.5$ as a proxy for the McGaugh IF is excluded at $>7\sigma$ by these results. Our determined value of $a_0$ is in good agreement with the commonly known value of $a_0=1.2$ or $1.3$. Although we do not show it here, a similar result is obtained for $\lambda$ of the $f_\lambda$ IF (Equation~(\ref{eq:IFlam})).

\begin{figure}
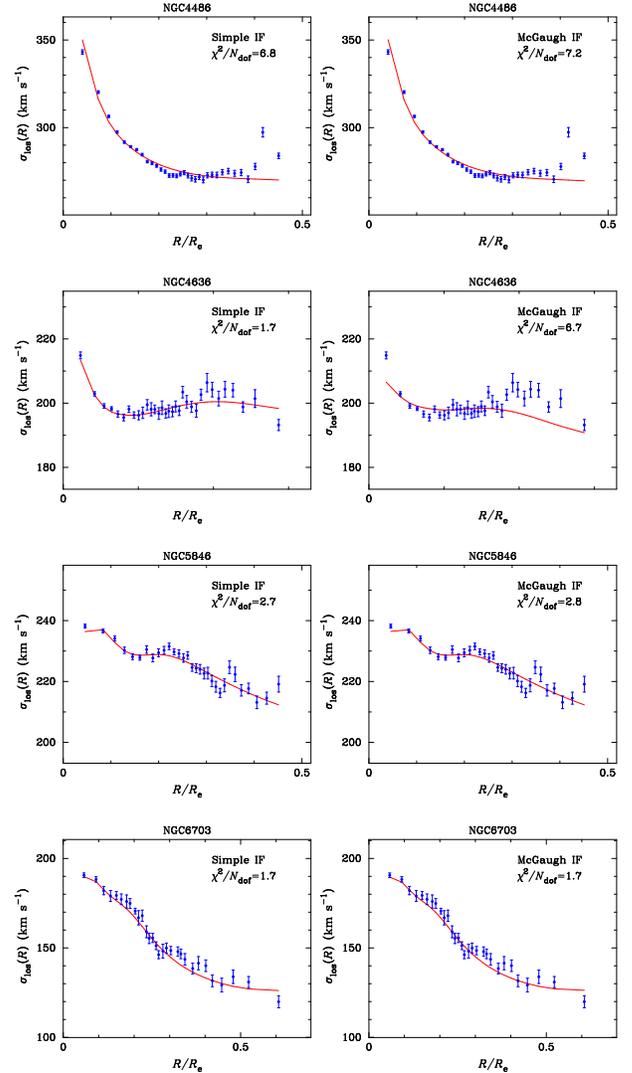
 %
  \centering
  \includegraphics[angle=-90,origin=c,scale=0.19]{Fig_VP_NGC4486_Spl.eps}
  \includegraphics[angle=-90,origin=c,scale=0.19]{Fig_VP_NGC4486_McG.eps}
  \includegraphics[angle=-90,origin=c,scale=0.19]{Fig_VP_NGC4636_Spl.eps}
  \includegraphics[angle=-90,origin=c,scale=0.19]{Fig_VP_NGC4636_McG.eps}
  \includegraphics[angle=-90,origin=c,scale=0.19]{Fig_VP_NGC5846_Spl.eps}
  \includegraphics[angle=-90,origin=c,scale=0.19]{Fig_VP_NGC5846_McG.eps}
  \includegraphics[angle=-90,origin=c,scale=0.19]{Fig_VP_NGC6703_Spl.eps}
  \includegraphics[angle=-90,origin=c,scale=0.19]{Fig_VP_NGC6703_McG.eps}
  \caption{Observed and best-fit \siglos\ profiles associated with the Simple and McGaugh IFs with $a_0=1.2\times 10^{-10}$~m~s$^{-2}$, for 4 ``spherical'' galaxies (ellipticities $\varepsilon < 0.1$).}
 \label{chi2test}
\end{figure}

\subsection{Eliminating models by direct fits to \siglos} \label{sec:chi2}
So far we have made Bayesian inferences about MOND IFs based on Monte-Carlo sampling of individual RARs and the stacked RAR.  However, our MC sampling uses the likelihood function which depends on $\chi^2$ of Equation~(\ref{eq:chisq}).  So it is interesting to ask if the RAR step is necessary: if simply fitting to \siglos\ is sufficiently discriminating.  

Figure~\ref{VDgNFW} already shows the best-fitting cases for the $\Lambda$CDM models with a gNFW halo.  Here, we consider similar checks within the MOND paradigm, for the Simple and McGaugh IFs with the fiducial value of $a_0$.  To illustrate, of the 24 \atl\ galaxies we only select galaxies having ellipticities $\varepsilon < 0.1$: they are NGC~4486, NGC~4636, NGC~5846, and NGC~6703, and are all slow rotators (see Table~1 of \citealt{CBS19} and references therein).

As shown in Figure~\ref{chi2test}, for NGC~5846 and NGC~6703 (bottom two sets of panels), both the IFs reproduce the observed velocity dispersion profiles reasonably well:  $\bar{\chi}^2_{\rm min}\lesssim 2.5$.  However, the McGaugh IF fails to provide a good fit to NGC~4636 (second from top, right); it has $\chi^2/N_{\rm dof}=6.7$.  At face value, this galaxy alone rules out the McGaugh IF as a viable MOND model.  (Although we do not show it here, allowing a different $a_0$ does not improve $\chi^2$ significantly.) For NGC~4486 (top), neither of the IFs provides an acceptable fit. (In contrast, Figure~\ref{VDgNFW} shows that the gNFW does fit well.) Again, at face value, \siglos\ of this one galaxy invalidates both the Simple and McGaugh functions as viable MOND IFs. Considering the significance of this result, it is important to investigate more spherical galaxies selected from large on-going and future IFS surveys such as MaNGA (see Section~\ref{sec:dis}). 

\section{Systematic Errors?} \label{sec:sys}
When inferring the ratio $a_{\rm X}/a_{\rm B}$ in the super-critical acceleration regime we marginalized over the $M_\star/L$ amplitude $\Upsilon_{\star 0}$ and the gradient strength parameter $K$ in Equation~(\ref{eq:MLgrad}) as well as the DM/PM profile parameters. We have considered a reasonable range of functional forms for the IF or DM profile (Equations~(\ref{eq:IFnu}), (\ref{eq:IFlam}), (\ref{eq:IFgam}), and (\ref{eq:gNFW})) which leave little possibility of hidden systematic error associated with the chosen functional forms of the IF or DM profile. However, is it possible that our treatment of the $M_\star/L$ gradient, Equation~(\ref{eq:MLgrad}) with $A=2.33$ and $B=6$, is too restrictive?  Whereas the ratio $A/B$ determines the radius within which the gradient matters, the product $KA$ determines $(M_\star/L)_{R=0}/(M_\star/L)_{\rm outer}$,
Since we already consider $0\le K< 1.5$, there is no need to consider varying $A$ as well.  Therefore, we have studied how our results change if we change $B$ (recall that $A/B=0.4$ is indicated by current data). 

\begin{figure}
  \vspace{-1em}
  \hspace{-5ex}
  \includegraphics[angle=0,origin=c,scale=0.5]{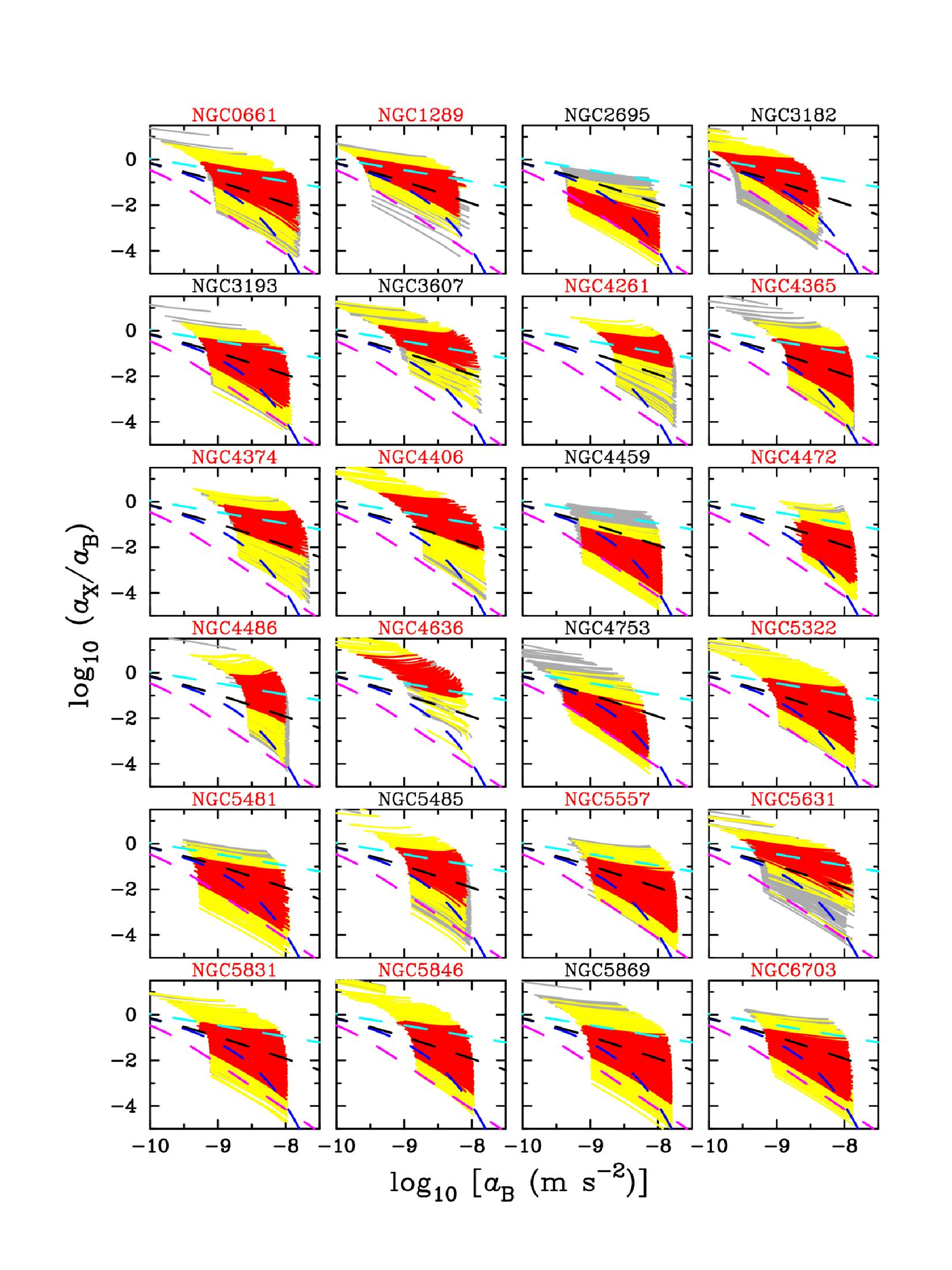}
  \vspace{-3em}
\caption{Results with artificially chosen value of $B=3$ in Equation~(\ref{eq:MLgrad}) which allows a smoother radial gradient over a larger radial range up to $0.8 R_{\rm e}$ than the \citet{vD17} limit of $0.4 R_{\rm e}$. The results on the ratio $a_{\rm X}/a_{\rm B}$ are little changed compared with those shown in Figure~\ref{RARgNFW}.}
\label{RARalt}
\end{figure}

Setting $B=3$ or $9$ ($R_{\rm lim}\approx 0.8R_{\rm e}$ or $0.25R_{\rm e}$; recall that current data prefer $R_{\rm lim}\approx 0.4R_{\rm e}$) makes little difference to the inferred ratio $a_{\rm X}/a_{\rm B}$, although the posterior PDFs of $K$ are shifted.  (The limiting cases of $B \rightarrow 0$ or $\infty$ would be equivalent to the no gradient case.)
Figure~\ref{RARalt} shows results with $B=3$. Thus, our results on $a_{\rm X}/a_{\rm B}$ are robust with respect to the choice of the gradient model unless it is dramatically different from Equation~(\ref{eq:MLgrad}).

Is it still possible that our finding that the Simple IF (Equation~\ref{eq:IFnu} with $\nu=1$) is preferred, is systematically in error?
Our numerical experiments show that McGaugh's IF could be preferred if we assume {\it no} $M_\star/L$ gradient ($K=0$) in any galaxies {\it and} the MOND IF given by Equation~(\ref{eq:IFlam}) (or the \citet{Ein} DM profile) is assumed.  On the other hand, if all elliptical galaxies have very strong $M_\star/L$ gradients ($K \gtrsim 1.2$), then Bekenstein's IF would be preferred. However, as we discuss below, we believe these extreme cases are unlikely.

As we discuss in the Introduction, no gradient at all ($K=0$) for any elliptical galaxies is not supported by a host of current observational studies. Our own Bayesian inferences \citep{CBS18} also do not support no gradient, although there are some cases where a weak gradient appears likely.  Very strong gradients ($K \gtrsim 1.2$) for {\it all} elliptical galaxies are also unlikely: $K=1$ is already on the highest side of the spectrum of current observational results.

\section{Discussion} \label{sec:dis}

We have obtained an empirical RAR  $a_{\rm X}/a_{\rm B} \approx 10^{-1} a_{+1} /a_{\rm B}$ around $a_{+1}\equiv 10^{+1} a_0$ in the super-critical acceleration regime $a_0 < a_{\rm B}<10^2 a_0$ through Bayesian inferences based on a host of MC sets of nearly spherical galaxies. This estimate, as unbiased result as possible, can narrow down DM/PM phenomenologies in galactic astrophysics and hence theoretical models of DM and modified dynamics or gravity. Verlinde's emergent gravity, Bekenstein's IF and the McGaugh et al.\ proposal of the exponentially decaying RAR (the Planck-like function) in the super-critical regime up to $100 a_0$ are all inconsistent at $>4\sigma$ with our results. For example, for McGaugh's function  to be valid in the super-critical acceleration regime, we must assume constant $M_\star/L$ in the central regions of {\it all} elliptical galaxies with the exponential IF (Equation~(\ref{eq:IFlam})) (or, the \citet{Ein} DM profile). At the other extreme, for Bekenstein's theory and Verlinde's theory to be valid in the super-critical acceleration regime, we must assume {\it all} ellipticals have stronger $M_\star/L$ gradients than the strongest reported to date.

A direct $\chi^2$ test of the McGaugh IF using the spherical galaxy NGC~4636 also rules out it as shown in Figure~\ref{chi2test}. Our results clearly support an RAR that can be described by the Simple IF as a mean (stacked) property. This result may well contain important information about DM properties or the property of gravity. Interestingly, the recently proposed scenario of baryon-DM interactions by \citet{FKP} is fully consistent with our empirical RAR.

In this context, it is worth noting that our present study is not a rigorous test of whether a universal RAR exists in elliptical galaxies, nor whether a universal acceleration scale $a_0$ exists, although we have investigated individual PDFs of MOND IF profiles (see Section~\ref{sec:chi2}). Nor do we have a particular paradigm in mind. Rather we have tried to extract the most information on DM or MG from the data. The on-going debate on the issue of the presence/absence of a universal RAR in rotating galaxies (\citealt{Rod18,McG18,Kro18}) is also relevant for dispersion-dominated elliptical galaxies. If a universal RAR truly exists, then both rotating galaxies and non-rotating galaxies must imply the same relation.

Our empirical RAR in the super-critical regime is not matched by the MUGS simulations of galaxy formation and evolution within the $\Lambda$CDM paradigm (Figure~\ref{PDFdm}). Since the other inputs to our $\Lambda$CDM estimate of the RAR (stellar to halo mass ratios, halo profiles and concentrations) are standard, the disagreement suggests that our empirical RAR provides a useful constraint on galaxy formation models within the context of $\Lambda$CDM.

While our present exploratory analysis of the super-critical acceleration regime of elliptical galaxies demonstrates the potential of using elliptical galaxies in addressing the DM problem, it also highlights the importance of quantifying $M_\star/L$ gradients. Large IFS samples, such as those provided by the MaNGA survey \citep{Bun15}, will allow a better determination of both \siglos\ and $M_\star/L$ gradients.  So we hope our exploratory analysis motivates a more precise determination of the RAR in these larger samples.  

\vspace{0.3in}

\acknowledgements
 We thank J. Khoury and M. van Putten for interesting and stimulating discussions. We thank them and M. Trodden for comments on the draft. We thank M. Cappellari for useful communications regarding the ATLAS$^{\rm 3D}$ results. We also would like to thank A. Kosowsky for constructive criticisms and an anonymous referee for insightful comments that helped us improve the analysis, presentation and discussion. This work was initiated at the University of Pennsylvania while KHC was on sabbatical leave in 2017. This research was supported by Basic Science Research Program through the National Research Foundation of Korea (NRF) funded by the Ministry of Education (NRF-2016R1D1A1B03935804). MB thanks NSF AST/1816330 for support.

\appendix

\section{Comments on Monte Carlo Sampling: SMC vs MCMC} \label{sec:mc2mc}

All the results in the main text use sets of parameters generated through Monte Carlo (MC) sampling algorithms. We have considered two algorithms: the SMC (our own algorithm written by K.-H. C.) and the MCMC implemented in the `emcee' package \citep{FM13}.  The SMC produces MC models that satisfy a certain criterion ($\bar{\chi}^2< \bar{\chi}^2_{\rm crit}$) without any correlation between successive iterations. Consequently, this method is slow but searches the entire parameter space within the prior ranges. Theoretically, it is based on a top-hat likelihood function, but the resulting posterior PDF of a parameter is not a top-hat shape because the PDF of $\bar{\chi}^2$ in the accepted parameter space does not follow a top-hat shape. Although this method is rather crude, tests using mock velocity dispersion profiles (see Section~\ref{sec:veri}) indicate that it is robust and satisfies the expected statistics reasonably (but conservatively rather than accurately).

\begin{figure}[b] 
\renewcommand\thefigure{A1}
 \centering

 \hspace{-1em}
 
  \vspace{-2em}
  \includegraphics[angle=0,origin=c,scale=0.48]{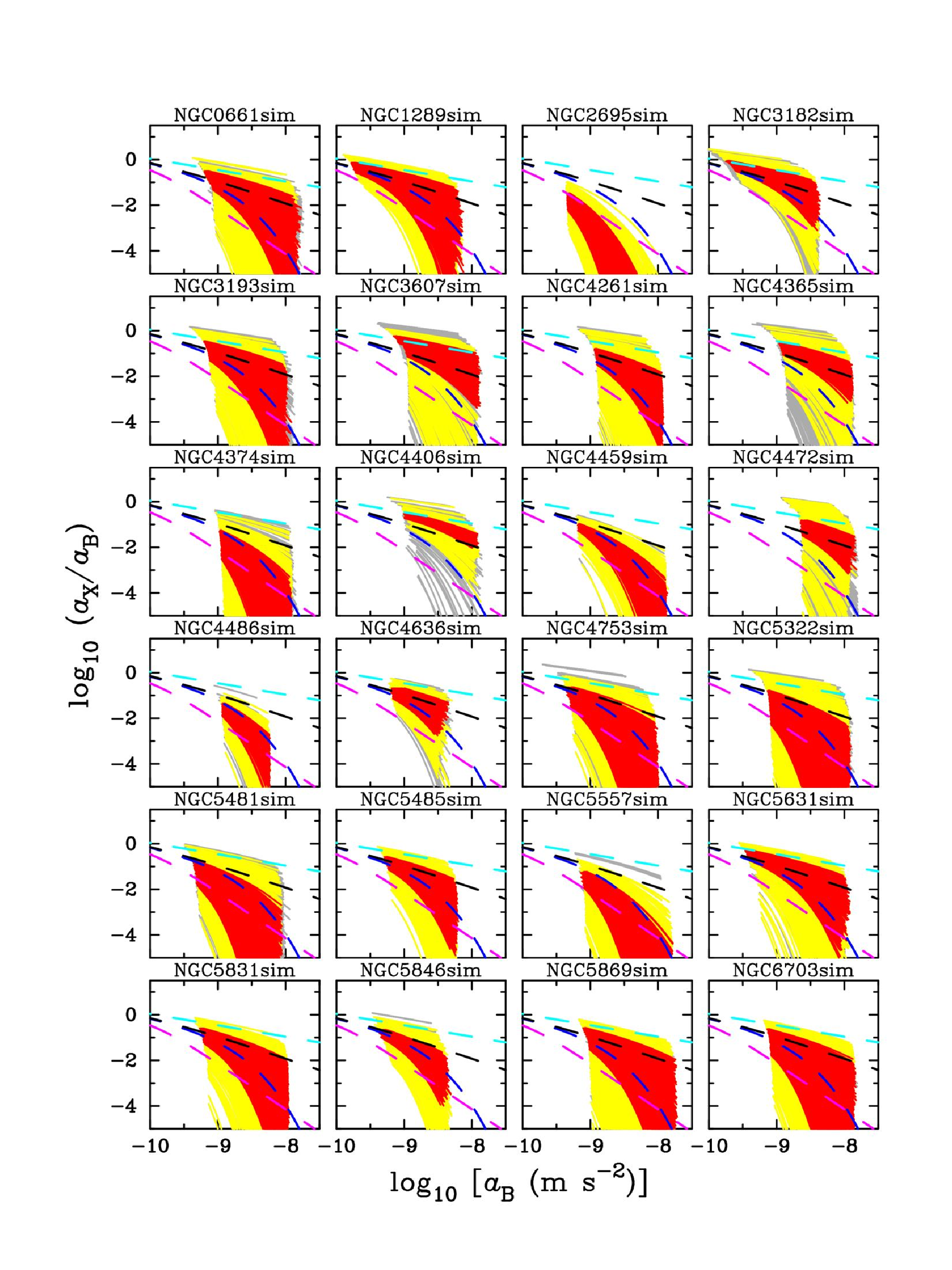}
  
  \vspace{-1em}
  \caption{Same as Figure~\ref{RARsim} but through the MCMC sampler.}
 \label{RARemcee}
\end{figure}

The MCMC method produces correlated MC models through self-proposed covariance matrices between iterations after a sufficient number of initial iterations. The algorithm is efficient and sophisticated, and hence, popular. We have tested it with a Gaussian likelihood (Equation~(\ref{eq:L})) using the same mock velocity dispersion profiles of Section~\ref{sec:veri}.  Figure~\ref{RARemcee} shows the analogue of Figure~\ref{RARsim}.  Compared with the SMC, the MCMC returns narrower distributions of $a_{\rm X}/a_{\rm B}$.  However, in some cases the posterior distributions actually rule out the input model (in this case, the McGaugh IF)! This is particularly so for NGC2695sim. This indicates that emcee with the Gaussian likelihood can sometimes return unrealistically narrow (over-confident) results.  This difficulty did not arise with the SMC, which is why we showed both SMC and MCMC sampling results in the main text.

\newpage

\section{Correlation of Parameters from the MCMC Sampling in the $\Lambda$CDM Case} \label{sec:corner}

We provide examples of the correlation of parameters from the MCMC sampling of models in the $\Lambda$CDM case. For this purpose, we chose the four roundest galaxies that can be modeled most reliably using spherical models: NGC 4486, 4636, 5846, and 6703. Figure~\ref{corner} exhibits the results.

\begin{figure}[b] 
\renewcommand\thefigure{A2}
 \centering
  \hspace{-1em}
  \includegraphics[angle=0,origin=c,scale=0.5]{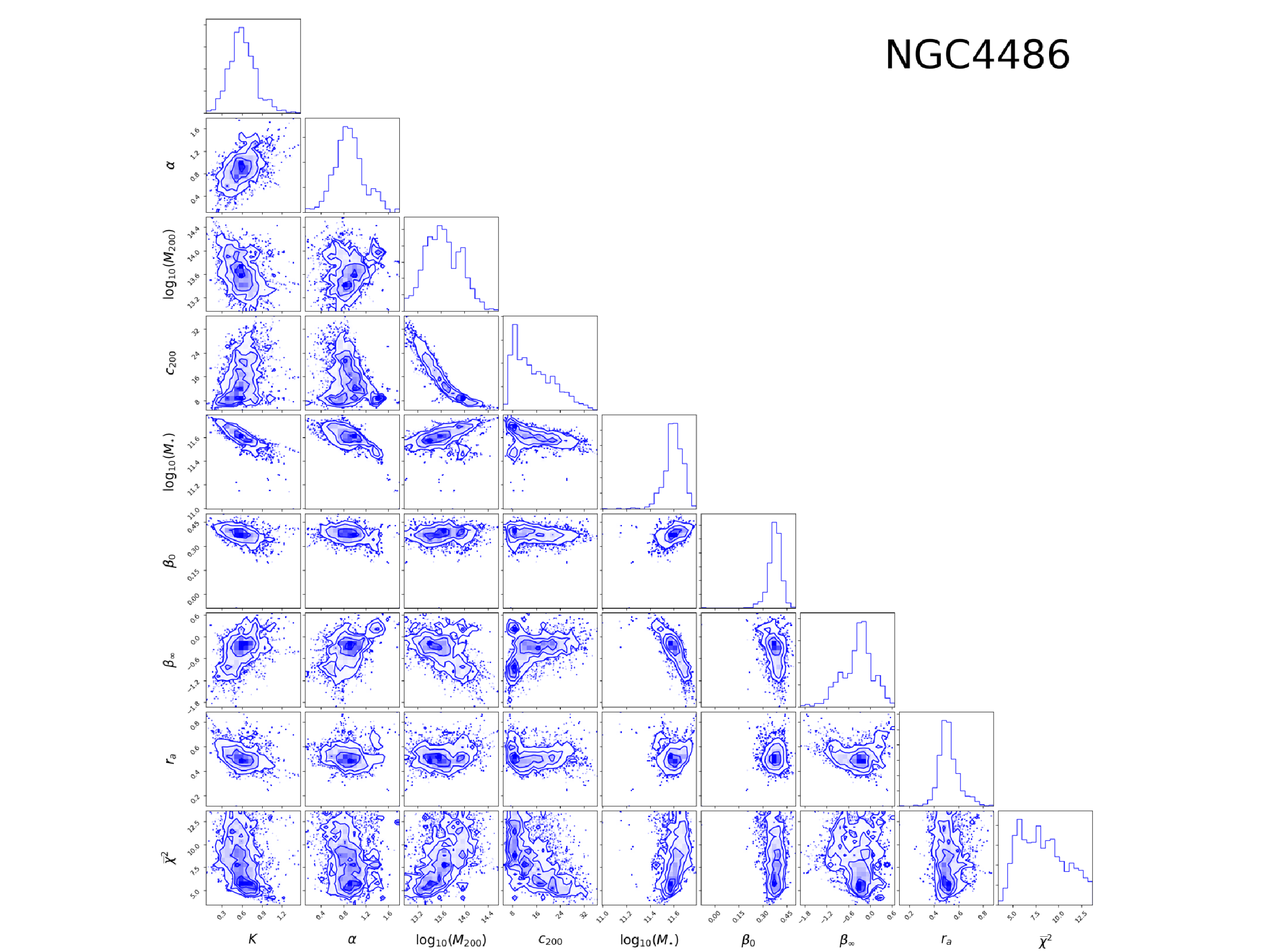}
  
  \vspace{1em}
  \includegraphics[angle=0,origin=c,scale=0.5]{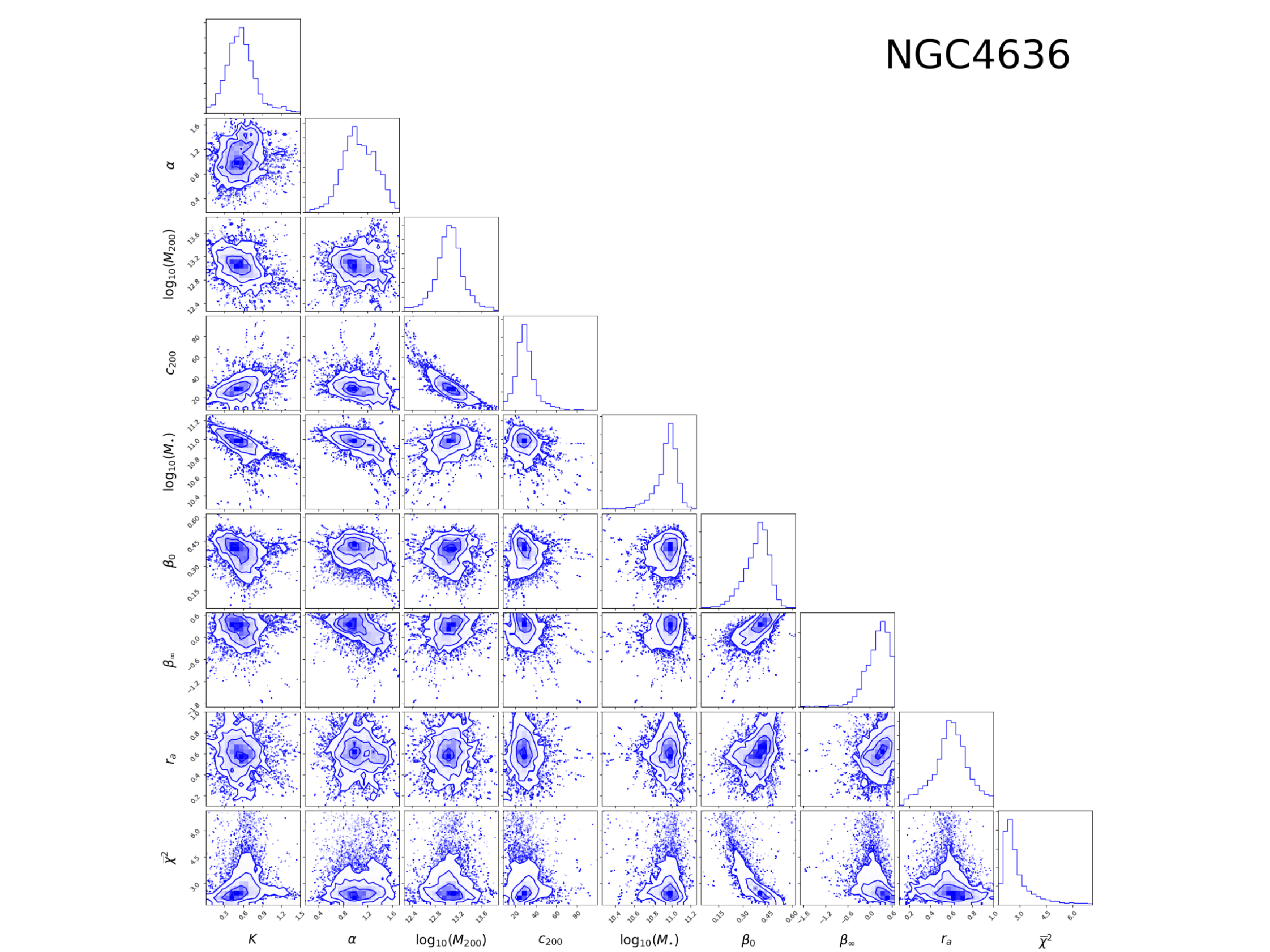}
  
  \caption{Correlation of parameters (Table~\ref{tab:prmt}) in the $\Lambda$CDM case through the MCMC sampler. Here stellar mass $M_\star$ refers to $\Upsilon_{\star 0} \times L_{r,{\rm MGE}}$ where $L_{r,{\rm MGE}}$ is the luminosity of the MGE light distribution in the SDSS $r$-band.}
 \label{corner}
\end{figure}

\begin{figure}[b] 
\renewcommand\thefigure{A2}
 \centering
  \hspace{-1em}
  \includegraphics[angle=0,origin=c,scale=0.5]{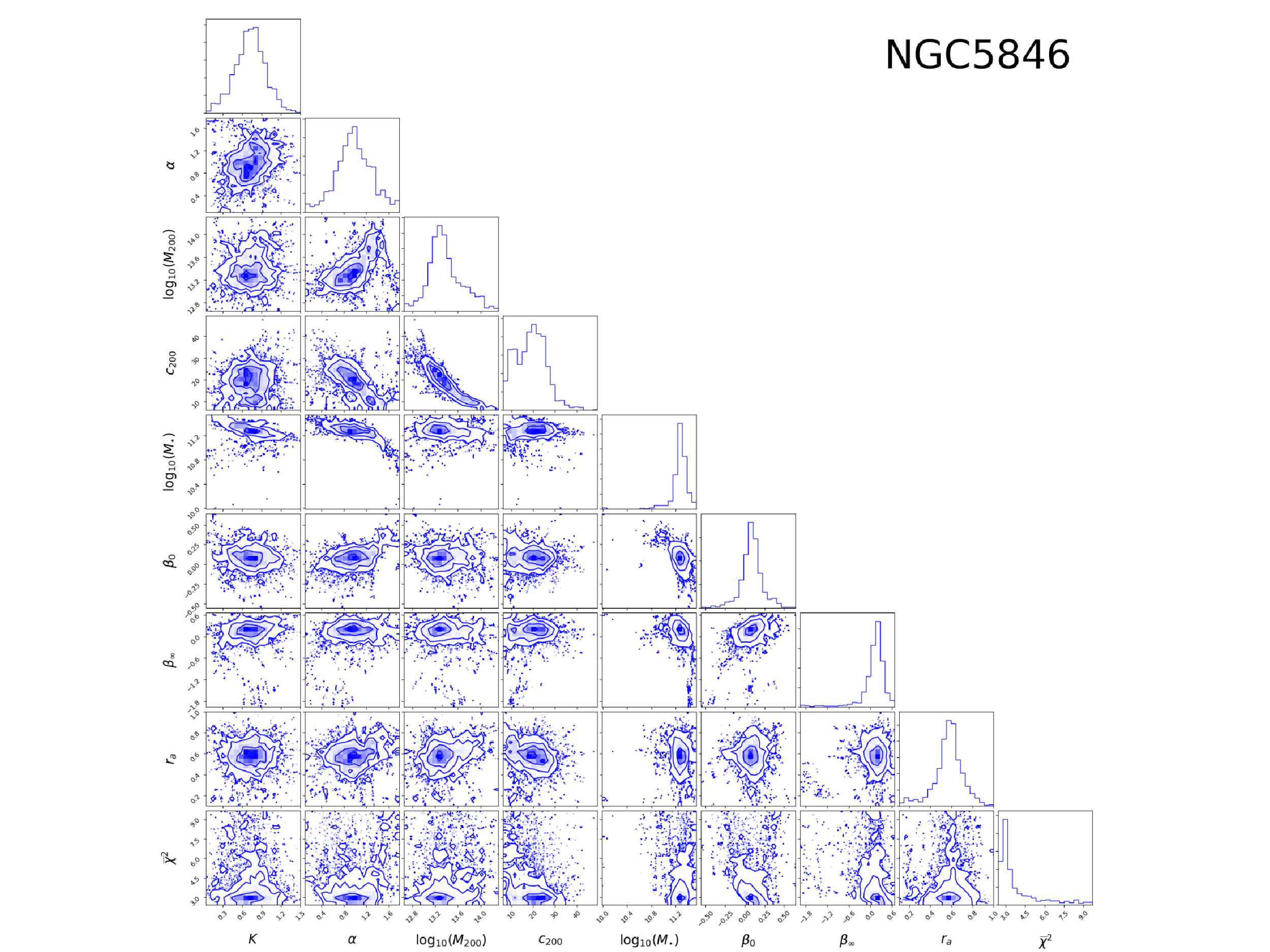}
  
  \vspace{1em}
  \includegraphics[angle=0,origin=c,scale=0.5]{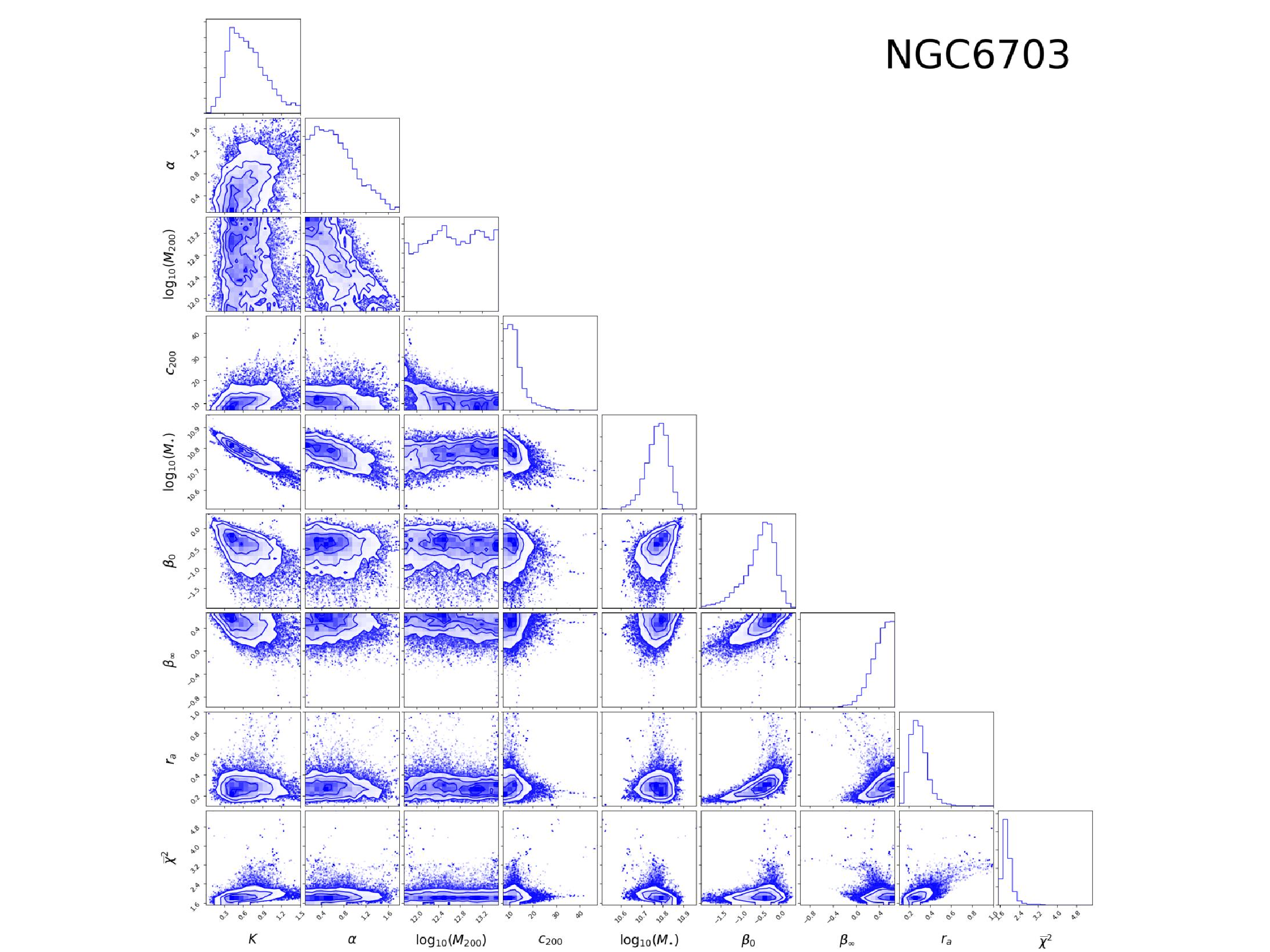}
  
  \caption{(Continued)}
 \label{corner}
\end{figure}


\begin{thebibliography}{}

\bibitem[Abazajian et al.(2009)]{DR7} Abazajian, K. N., Adelman-McCarthy, J.~K., Ag\"{u}eros, M.~A., et al.\ 2009, \apjs, 182, 543

\bibitem[Allison \& Dunkley(2014)]{AD14} Allison, R., Dunkley, J. 2014, \mnras, 437, 3918
    
\bibitem[Alton et al.(2017)]{Alt17} Alton, P. D., Smith, R. J., Lucey, J. R. 2017, \mnras, 468, 1594

\bibitem[Alton et al.(2018)]{Alt18} Alton, P. D., Smith, R. J., Lucey, J. R. 2018, \mnras, 478, 4464

\bibitem[Bekenstein(2004)]{Bek} Bekenstein,  J. D. 2004, \prd, 70, 083509

\bibitem[Berezhiani \& Khoury(2015)]{Kho} Berezhiani,  L., Khoury,  J. 2015, \prd, 92, 103510

\bibitem[Bernardi et al.(2018)]{Ber18} Bernardi, M., Sheth, R. K., Dominguez-Sanchez, H., et al.\ 2018, MNRAS, 477, 2560

\bibitem[Bertone \& Tait(2018)]{BT18} Bertone, G., Tait, T. M. P. 2018, \nat, 562, 51 
  
\bibitem[Binney \& Tremaine(2008)]{BT} Binney, J., Tremaine, S. 2008, Galactic Dynamics, (2nd ed.; Princeton, NJ: Princeton Univ.\ Press)

\bibitem[Blanchet \& Le Tiec(2009)]{BL} Blanchet, L.,  Le Tiec, A. 2009, \prd, 80, 023524
  
\bibitem[Bundy et al.(2015)]{Bun15} Bundy, K., et al.\ 2015, \apj, 798, 7

\bibitem[Burrage et al.(2017)]{BCM} Burrage, C., Copeland, E. J., Millington,  P. 2017, \prd, 95, 064050  
  
\bibitem[Cappellari et al.(2011)]{Cap11} Cappellari, M., Emsellem, E., Krajnovi\'{c}, D., et al.\ 2011, \mnras, 413, 813
    
\bibitem[Chae, Bernardi \& Kravtsov(2014)]{CBK14} Chae, K.-H., Bernardi, M., Kravtsov,  A. V. 2014, \mnras, 437, 3670

\bibitem[Chae, Bernardi \& Sheth(2018)]{CBS18} Chae, K.-H., Bernardi, M., Sheth,  R. K. 2018, \apj, 860, 81

\bibitem[Chae, Bernardi \& Sheth(2019)]{CBS19} Chae, K.-H., Bernardi, M., Sheth,  R. K. 2019, \apj, 874, 41

\bibitem[Chae \& Gong(2015)]{CG15} Chae, K.-H., Gong, I.-T. 2015, \mnras, 451, 1719 

\bibitem[Davis \& McDermid(2017)]{DM17} Davis, T. A., McDermid, R. M. 2017, \mnras, 464, 453
  
\bibitem[Desmond(2017)]{Des} Desmond, H. 2017, \mnras, 464, 4160
  
\bibitem[Diemer \& Kravtsov(2015)]{DK15} Diemer, B., Kravtsov, A. V. 2015, \apj, 799, 108

\bibitem[Einasto(1965)]{Ein} Einasto, J. 1965, TrAlm, 5, 87  

\bibitem[Famaey \& Binney(2005)]{FB} Famaey, B., Binney, J. 2005, \mnras, 363, 603

\bibitem[Famaey, Khoury \& Penco(2018)]{FKP} Famaey, B., Khoury, J., Penco, R. 2018, JCAP, 03, 038

\bibitem[Famaey \& McGaugh(2012)]{FM} Famaey, B., McGaugh, S. S. 2012, LRR, 15, 10

\bibitem[Foreman-Mackey et al.(2013)]{FM13} Foreman-Mackey, D., Hogg, D. W., Lang, D., Goodman, J. 2013, \pasp, 125, 306 
  
\bibitem[Gerhard et al.(2001)]{Ger01} Gerhard, O., Kronawitter, A., Saglia, R. P., Bender, R. 2001, \aj, 121, 1936

\bibitem[Janz et al.(2016)]{Jan16} Janz, J., Cappellari, M., Romanowsky, A. J.,  Ciotti, L., Alabi, A., Forbes, D. A. 2016, \mnras, 461, 2367  
  
\bibitem[Keller \& Wadsley(2017)]{KW} Keller, B. W., Wadsley, J. W. 2017, \apjl, 835, 17 
  
\bibitem[Kent(1987)]{Ken} Kent, S. M. 1987, \aj, 93, 816

\bibitem[Kroupa(2002)]{Kro02} Kroupa, P. 2002, Science, 295, 82
  
\bibitem[Kroupa et al.(2018)]{Kro18} Kroupa, P., Banik, I., Haghi, H., Zonoozi, A., et al. 2018, NatAs, in press  

\bibitem[La Barbera et al.(2016)]{LaB16} La Barbera, F., Vazdekis, A., Ferreras, I., et al., 2016, \mnras, 457, 1468
  
\bibitem[Lelli et al.(2017)]{Lel} Lelli, F., McGaugh, S. S., Schombert, J. M., Pawlowski, M. S. 2017, \apj, 836, 152

\bibitem[Li et al.(2018)]{Li18} Li, P., Lelli, F., McGaugh, S., Schombert, J. 2018, \aap, 615, 70  
  
\bibitem[Ludlow et al.(2017)]{Lud} Ludlow,  A. D., Benitez-Llambay, A., Schaller, M., Theuns, T., et al.\ 2017, \prl, 118, 1103

\bibitem[Mandelbaum, Seljak \& Hirata(2008)]{Man08} Mandelbaum, R., Seljak, U., Hirata, C. M. 2008, JCAP, 08, 006

\bibitem[Mandelbaum et al.(2016)]{Man16} Mandelbaum, R., Wang, W., Zu, Y., et al.\ 2016, \mnras, 457, 3200 

\bibitem[Mart\'{i}n-Navarro et al.(2015)]{MN} Mart\'{i}n-Navarro, I., La Barbera, F., Vazdekis, A., Falc\'{o}n-Barroso, J., Ferreras, I. 2015, \mnras, 447, 1033

\bibitem[McGaugh(2004)]{McG04} McGaugh, S. S. 2004, \apj, 609, 652

\bibitem[McGaugh(2008)]{McG08} McGaugh, S. 2008, \apj, 683, 137

\bibitem[McGaugh, Lelli \& Schombert(2016)]{MLS} McGaugh, S. S., Lelli, F., Schombert, J. M. 2016, \prl, 117, 201101

\bibitem[McGaugh et al.(2018)]{McG18} McGaugh, S. S., Li, P., Lelli, F., Schombert, J. M. 2018, NatAs, in press

\bibitem[Meert, Vikram \& Bernardi(2015)]{Mee} Meert, A., Vikram, V., Bernardi, M. 2015, \mnras, 446, 3943

\bibitem[Milgrom(1983)]{Mil} Milgrom, M. 1983, \apj, 270, 371

\bibitem[Milgrom(2012)]{Mil12} Milgrom, M. 2012, \prl, 109, 131101 

\bibitem[Navarro et al.(2017)]{Nav} Navarro, J. F., Ben\'{i}tez-Llambay, A., Fattahi, A., Frenk, C. S., et al.\ 2017, \mnras, 471, 1841
    
\bibitem[Navarro, Frenk \& White(1997)]{NFW} Navarro, J. F., Frenk, C. S., White, S. D. M. 1997, \apj, 490, 493
  
\bibitem[Oldham \& Auger(2018)]{Old18} Oldham, L., Auger, M. 2018, \mnras, 474, 4169  

\bibitem[Rodrigues et al.(2018)]{Rod18} Rodrigues, D. C., Marra, V., del Popolo, A., Davari, Z. 2018, NatAs, 2, 668

\bibitem[Sanders \& Noordermeer(2007)]{SN07} Sanders, R.~H., Noordermeer, E. 2007, \mnras, 379, 702

\bibitem[Sarzi et al.(2018)]{Sar18} Sarzi, M., Spiniello, C., La Barbera, F., Krajnovi\'{c}, D., van den Bosch, R. 2018, \mnras, 478, 4084
  
\bibitem[Shankar et al.(2017)]{Shan17}  Shankar, F., et al.\ 2017, \apj, 840, 34
  
\bibitem[Sonnenfeld et al.(2015)]{Son15} Sonnenfeld, A., Treu, T., Marshall, P. J., Suyu, S. H., Gavazzi, R., Auger, M. W., Nipoti, C. 2015, \apj, 800, 94 

\bibitem[Sonnenfeld et al.(2018)]{Son18} Sonnenfeld, A., Leauthaud, A.,  Auger, M. W., et al.\ 2018, \mnras, 481, 164

\bibitem[Tenneti et al.(2018)]{Ten} Tenneti, A., Mao, Y.-Y., Croft, R. A. C. , Di Matteo, T., Kosowsky,  A., Zago, F., Zentner, A. R. 2018, \mnras, 474, 3125
  
\bibitem[van Dokkum et al.(2017)]{vD17} van Dokkum, P., Conroy, C., Villaume, A. , Brodie, J., Romanowsky, A. J. 2017, \apj, 841, 68 

\bibitem[van Putten(2018)]{vP18} van Putten, M. H. P. M. 2018, \mnras, 481L, 26
    
\bibitem[Verlinde(2017)]{Ver} Verlinde,  E. P. 2017, SciPost Physics, 2, 016 (arXiv:1611.02269)

\bibitem[Wall \& Jenkins(2012)]{WJ12} Wall, J. V., Jenkins, C. R. 2012, Practical Statistics for Astronomers, (2nd ed.; Cambridge Univ.\ Press)

\end{thebibliography}
\end{document}